\documentclass[12pt,review,authoryear]{elsarticle}
\usepackage{amsmath,amsfonts,amssymb,amscd,amsthm,amsbsy, color }

\usepackage{graphicx,epsfig}

\newcommand\beq[1]{ \begin{equation}\label{#1} }
\newcommand{\eeq}{ \end{equation} }
\newcommand{\beqno}{ \[ }
\newcommand{\eeqno}{ \] }
\newcommand\beqa[1]{ \begin{eqnarray} \label{#1}}
\newcommand{\eeqa}{ \end{eqnarray} }
\newcommand{\beqano}{ \begin{eqnarray*} }
\newcommand{\eeqano}{ \end{eqnarray*} }
\newcommand\equ[1]{{\rm (\ref{#1})}}

\newcommand{\red}{}

\newtheorem{theorem}{Theorem}

\newtheorem{remark}[theorem]{Remark}

\journal{Icarus}

\begin{document}

\begin{frontmatter}

\title{The effect of Poynting-Robertson drag on the triangular Lagrangian points}

\author[CL]{C. Lhotka\corref{cor1}\fnref{fn1}}
\ead{christoph.lhotka@oeaw.ac.at}

\author[AC]{A. Celletti}
\ead{celletti@mat.uniroma2.it}

\cortext[cor1]{Corresponding author}

\fntext[fn1]{Now working at the Space Research Institute, Austrian Academy 
of Science, Schmiedlstrasse 6, 8042 Graz, Austria}

\address[CL]{Institut f\"ur Astrophysik, 
Universit\"at Wien, T\"urkenschanzstra\ss e 17\\
A-1180 Wien (Austria);}

\address[AC]{Department of Mathematics, 
University of Roma Tor Vergata, Via della Ricerca Scientifica 1,
00133 Roma (Italy)}

\begin{abstract} We investigate the stability of motion close to the Lagrangian
equilibrium points $L_4$ and $L_5$ in the framework of the spatial, elliptic,
restricted three-body problem, subject to the radial component of
Poynting-Robertson drag.  For this reason we develop a simplified resonant
model, that is based on averaging theory, i.e.  averaged over the mean
\red{anomaly} of the perturbing planet.  We find \red{temporary stability of
particles displaying a tadpole motion in the 1:1 resonance.} From the linear
stability study of the averaged simplified resonant model, we find that the
time of temporary \red{stability} is proportional to $\beta\ a_1\ n_1$, where
$\beta$ is the ratio of the solar radiation over the gravitational force, and
$a_1$, $n_1$ are the semi-major axis and the mean motion of the perturbing
planet, respectively. We extend previous results (\cite{Mur94}) on the
asymmetry of the stability indices of $L_4$ and $L_5$ to a more realistic force
model. Our analytical results are supported by means of numerical simulations.
We implement our study to Jupiter-like perturbing planets, that are also found
in extra-solar planetary systems.  \end{abstract}



\end{frontmatter}



\section{Introduction}

The motivation of our study is to understand better the effect of stellar
radiation on resonant interactions between the motion of dust and
a planet in planetary systems. The subject has already been
studied in \cite{Der94}, where the authors propose that dust may
be transported from the main belt to the Earth by temporary
resonant capture; the trapping mechanism for exterior mean motion
resonances (MMRs) has been studied in full detail in \cite{Bea94,
Wei93}. Only outer resonances have been found to be stable
(see \cite{Sic93} and references therein); the effect of drag on
motions close to the Lagrange points is treated in \cite{Mur94},
where the author finds asymmetric stability indices for the
triangular points. The three-dimensional orbital evolution of
dust particles is also studied in \cite{Lio97, Lio95}.
\cite{Kor13} has demonstrated a trapping mechanism of dust
particles in Earth's quasi-satellite resonance. The literature on
the subject is wide and we refer the reader to the
bibliography\footnote{Among all papers on the subject, let us
quote the following results. In \cite{Pas09a} the authors
investigate the eccentricity evolution of particles under the
effect of non-radial wind, while the interplay with MMRs has been
treated in \cite{Pas09b}. Stability times have been derived in
\cite{Kla08a}, the effect of the non-radial components of solar
wind on the motion of dust near MMRs is treated in \cite{Kla08b},
non-radial dust grains close to MMRs are subject of \cite{Koc08}.
The dynamical effect of Mars on asteroidal dust particles has been
investigated in \cite{Esp08}, the resonance with Neptune has been
treated in \cite{Koc12}. The triangular libration points have also
been treated in \cite{Sin14}, the collinear ones in \cite{Ste08}.
Out of plane equilibrium points have been found in \cite{Das08}.
To this end, a critical  review of the PR effect, that has been
used in the literature - \cite{Buretal79} - can be found in
\cite{Kla14} with a justification of the authors in \cite{Bur14}.
}. \\

The goals of this paper are the following: i) most analytical
studies mentioned above are based on the circular or/and planar,
restricted, three-body problem; we therefore aim to extend these
results to the case of the spatial, elliptic, restricted
three-body problem (SERTBP); this is of particular importance,
since stellar radiation forces act in 3D space. ii) The 1:1 MMR
under Poynting-Robertson (hereafter PR) effect is only poorly studied by analytical means (with some
exception found in \cite{Mur94} - but based on the circular
problem); the reason can maybe found in the fact that standard
expansions of the perturbing function do not converge if the ratio
of the semi-major axes of the perturber and dust particle tends to
unity; we therefore aim to use the equilateral perturbing function
instead of the standard one to properly treat the case of the 1:1
MMR. iii) It is commonly accepted that Poynting-Robertson drag
destabilizes inner resonances with an external perturber, while
temporary capture may be found for outer resonances with an
internal perturber (\cite{Bea94}); we therefore aim to provide with this
study the missing link with the co-orbital resonant regime
of motion.\\

There are different kinds of forces that need to be taken into account to model
the dynamics of interplanetary dust, that strongly depend on the size of the
particles of interest (\cite{Gus94}): solar gravity, stellar radiation
pressure, the Lorentz force, planetary perturbations, PR drag,
and solar wind drag. In our study we concentrate on particles within the size
range \red{from $1\ \mu m$ to $200\ \mu m$}, where the Lorentz force can be
safely neglected, the primary force is solar gravity and stellar radiation
pressure, second order effects are additional planetary perturbations,
about the same order of the PR effect. Precisely, we are interested in the interplay
of these second order effects on motion of dust particles that are situated
inside the 1:1 mean motion resonance with a planet.  As we will show the PR
effect does not only strongly influence the orbital life-time of particles, but
also the location of the resonance in the orbital element space of the
interplanetary dust particle.  \\

The models describing the three-body problem are introduced in
order of increasing difficulty: from the circular-planar case to
the elliptic-inclined one. The models include the effect of
Poynting-Robertson drag. Three case studies are identified:
Jupiter and two samples, which are representative of extrasolar
planetary systems. Using the equations of motion averaged over the
mean anomaly, we are able to detect stationary solutions and to
describe them in terms of the parameters of the system. We also
add a discussion about the eigenvalues of the linearized vector
field as a function of the dissipative parameter. We conclude wih
a comparison with the unaveraged vector field. \\

The content of this paper is the following. In
Section~\ref{MatMod} we define the mathematical model and the
equations of motion that we use for our study. In
Section~\ref{ResMod} we derive a resonant model, valid close to
the 1:1 resonance that is based on averaging theory. We
investigate the equilibria of the averaged problem in the
framework of the planar, spatial, circular and elliptic restricted
three body problems in Section~\ref{StaSol}, and perform a linear
stability study of the SERTBP in Section~\ref{LinSta}. A numerical
survey based on the unaveraged equations of motions can be found
in Section~\ref{NumSur}.  A summary of our conclusions is given in
Section~\ref{SumCon}; supplementary calculations that may help the
reader to
reproduce our results can be found in the Appendices. \\

\section{Mathematical model and case studies}
\label{MatMod}

We investigate the dynamics of dust-size particles in the framework of the
spatial, elliptic restricted three-body problem (SERTBP), in which the central
body is the source of electromagnetic radiation, while the second largest body
does not radiate. We denote the celestial bodies involved in our model,
respectively as the \sl central body \rm (e.g., the star), the \sl secondary
\rm body (e.g., a planet), and the \sl third \rm body (e.g., a dust particle).
The third body is thus subject to two different kinds of forces, as described below.

\begin{itemize}

\item[1)]{\bf Gravitational Attraction (GA)}

Let $\vec r$, $\vec r_1$ be the vectors of the third and the secondary
from the central body in \red{a
coordinate system} with the origin coinciding with the central body.
We denote by $r=\|\vec r\|$, $r_1=\|\vec r_1\|$ the
distances of the third and the secondary from the origin, and by $\Delta$ the
mutual distance of the third and the secondary bodies. Let ${\mathcal G}$ be the gravitational constant,
and $m_0$, $m_1$, $m$ be the mass of the central, the secondary,
and the third body, respectively. In this setting, the gravitational force can be derived
from the force function
\beq{eq:F1}
U_{grav}=-\frac{{\mathcal G}(m_0+m)}{r}-
{\mathcal G} m_1 \big(\frac{1}{\Delta}-\frac{\vec r \cdot \vec {r}_1}{r_1^3}
\big) \ .
\eeq
As it is standard in the \sl restricted \rm three body problem,
we assume that the third body does not influence the motion of the other two
bodies, and formally we set $m=0$.

\item[2)]{\bf Solar Radiation (SR)}

Let $\hat{ r}=\vec r / r$ be the radial unit vector, $\dot r=dr/dt$ be the
radial velocity, and $\vec v=d{\vec r}/dt$ be the instantaneous velocity vector of
the \red{third mass}. We denote by the dimensionless parameter
$\beta$ the ratio of the radiation pressure force that is felt by a
particle of radius $r_p$ and density $\rho$, over the gravitational force of the
central body of mass $m_0$ at distance $r$.
Let $c$ be the speed of light and $\gamma\equiv 1+s_w$, where $s_w$ is the ratio of the
net force of solar wind over the net force due to the Poynting-Robertson effect. The forces of
interest, solar radiation pressure (hereafter SRP), Poynting-Robertson drag (denoted as PR), and solar
wind drag (hereafter SW), are given by (see, e.g., \cite{Buretal79, Bea94, Lio95}):
\beq{eq:F2}
\vec {F}_{EMF}=\frac{{\mathcal G}m_0}{r^2}\beta
\bigg(
\big(1-\gamma
\frac{\dot{r}}{c}\big)
\hat { r}-\gamma
\frac{\vec v}{c}
\bigg) \ .
\eeq
Notice that for $\beta=0$ there is no solar radiation, for $\gamma=0$ we just consider SRP, which reduces to a conservative effect,
while for $\gamma=1$ the effect of the solar wind is neglected. We also notice that we neglect higher order
terms, precisely the transversal component of solar wind drag (\cite{Kla13}). From now on, we shall focus only on
the case $\gamma=1$, which corresponds to PR drag without SW.
\red{This choice is motivated by the fact that PR effect is considered as the most
important non-gravitational effect acting on dust particles of the size we consider
in this paper (compare with \cite{Gru85}). However, SW will certainly deserve a
further study, since in \cite{Kla14b} it is shown that for non Maxwell-Boltzmann
velocity distributions of the solar wind, the SW effect is more important than the action of the
radiation, as for the secular orbital evolution.}

\end{itemize}

\subsection{Equations of motion in the Cartesian framework}

The equations of motion, in vector notation, for the massless particle
within the SERTBP under the effect of PR drag, can be easily derived
from \equ{eq:F1} and \equ{eq:F2}:
\beq{eq:vec}
\frac{d^2\vec{r}}{d t^2}=
-\mu_0 (1-\beta )\frac{\vec{r}}{r^{3}}-
\mu_1\left(
\frac{\vec{r}_1}{r_1{}^3}+
\frac{\vec{r}-\vec{r}_1}{\Delta ^3}
\right)-
\frac{\mu_0 \beta \gamma}{c r^2}\left(\dot{r} \hat{ r}+\vec{v}\right) \ ,
\eeq
where we have introduced the mass parameters \(\mu_0={\mathcal G} m_0\) and  $\mu_1={\mathcal G} m_1$. Let
us write
\beqano
\vec{r}=(x,y,z) \ ,\qquad
\vec{v}=\left(v_x,v_y,v_z\right) \ ,\qquad
\vec{r}_1=\left(x_1,y_1,z_1\right) \ ,
\eeqano
where $(x,y,z)$ denotes the position of the third body in the Cartesian space,
$(v_x,v_y, v_z)$ labels its velocity, and $(x_1,y_1,z_1)$ denotes the
position of the secondary in the Cartesian space. In this setting
the various quantities appearing in \equ{eq:vec} can be written in components as
\beqano
&&r=\|\vec r\|=\sqrt{x^2+y^2+z^2} \ , \quad
r_1=\|\vec{r}_1\|=
\sqrt{x_1^2+y_1^2+z_1^2} \ , \\
&&\Delta =\|\vec{r}-\vec{r}_1\|=
\sqrt{\left(x-x_1\right)^2+\left(y-y_1\right)^2+\left(z-z_1\right)^2} \ , \\
&&\hat{ r}=\left(\hat{x},\hat{y},\hat{z}\right)\equiv
(\frac{x}{r},\frac{y}{r},\frac{z}{r}) \ , \quad
\dot{r}=\frac{d r}{d t}=\frac{\vec{v}\cdot\vec r}{r}=
 \frac{v_xx+v_yy+v_zz}{r} \ .
\eeqano
The above expressions allow us to write the components of the electro-magnetic
force in \equ{eq:F2} in explicit form.

\subsection{Parameters and Units}\label{sec:parameters}
In the following discussion we simplify our problem by a proper choice of
the units of measure. Let ${\mathcal G}=1$, the unit of mass coincide
with $m_0+m_1+m$, the unit of length be the semi-major axis of the secondary
$a_1$. From  $\mu_*=\mu_0+\mu_1+{\mathcal G}m$ equal $1$, we can write
$\mu_1=1-\mu_0$, since we assumed $m=0$. From $n_1^2a_1^3=\mu_*$  and setting $a_1=1$, then
the mean motion of the secondary becomes $n_1=1$ and the revolution period
equals $2\pi$.
In our units the speed of
light\footnote{Converting the speed of light $c=299 792 458\ m/s$ in units $AU/d$, we obtain $c=172.672\, AU/d$;
setting $a_1=n_1=1$, we find $c=172.672(n_1a_1)^{-1}$ in the units in which the secondary is at distance
equal to unity and its period of revolution is $2\pi$.}
 is equal to $c=22946.5$ for the case of Sun-Jupiter, and
$c=10065.3$ in the case of Sun-Earth. We assume that the third
particle is spherical, and composed of silicates; we limit our
study to particles with radii ranging \red{from $1\ \mu m$ to $200\ \mu m$
- with values of $\beta$ ranging from $0$ to $0.1$
($\beta\simeq0.2/r_p$, see \cite{Bea94})}.

\subsection{Case studies}\label{sec:case}

\begin{table}
\caption{Case studies investigated in this work; additional parameters are specified in
Sections~\ref{sec:parameters} and \ref{sec:case}. The speed of light $c$ is given in units such
that the secondary is at distance 1 and its period of revolution is $2\pi$. }
$
\begin{array}{ccccc}
\hline\hline
id. & m_1[m_J] & a_1[AU] & P_1[days] & c \\
\hline
Jupiter (red) & 1 & 5.203 & 4344.68 & 22946.5 \\
 1 & 1.5 & 2 & 800 & 10992.6 \\
 2 & 0.6 & 0.05 & 5 & 2748.16 \\
\hline\hline
\end{array}
$
\label{t:sys}
\end{table}

\begin{figure}
\centering
\includegraphics[width=0.6\linewidth]{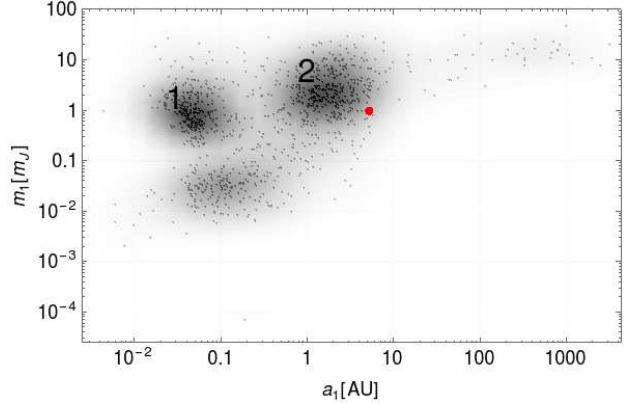}
\caption{Semi-major axis $a_1$ vs. mass ratio $m_1$ of 1796 exo-planets
(source www.exoplanets.eu). Dark regions, like those labeled 1 and 2,
indicate high density regimes; the red dot denotes Jupiter.}
\label{f:exo}
\end{figure}

Let $m_S$, $m_J$ be the mass of the Sun and Jupiter, respectively; let $P_1$
denote the period of revolution of the secondary. We are going to implement our
study on the Sun-Jupiter system, and two representative exo-planetary systems, that
are obtained as follows: in Figure~\ref{f:exo} we present the
data of 1796 known exo-planets in a
suitable $a_1 - m_1$ plot: the regions 1 and 2 define the two most dominant
high density regimes of mass and distance in the parameter space $(a_1,m_1)$.
The red dot represents Jupiter.  To obtain the proper periods $P_1$, to be
able to calculate $n_1$ and $c$ in our choice of units, we choose the closest
known exo-planets in our database to the centers of the regions 1, 2, that also have a central
star similar to our Sun: CoRoT-16 b ($a_1=0.061, m_1=0.53,
P=5.35$), and 16CygB b ($a_1=1.68, m_1=1.68, P=799.5$). The star CoRoT-16 has
spectral type G5V with mass $m_0=1.098m_S$, the star 16 Cyg B is of type G2.5V
with mass $m_0=1.01\, m_S$. Using the relation $n_1=2\pi/P_1$, we are yet able
to calculate $c=172.672(n_1a_1)^{-1}$ in proper units as summarized in
Table~\ref{t:sys}.

\section{Resonant variables and averaged equations of motion}
\label{ResMod}

Following \cite{BroSho64}, we write the potential\footnote{We use opposite signs
w.r.t. \cite{BroSho64}.}, taking the \red{star} as origin of the coordinates' frame:
\beq{UBS}
U_{grav}=-\frac{\mu_*}{r}-
\mu_1 \big(\frac{1}{\Delta}-\frac{1}{r}-\frac{\vec r \cdot \vec {r}_1}{r_1^3}\big) \ .
\eeq

Let us denote by $a$, $e$, $i$, $\omega$, $\Omega$, $M$ the standard orbital
elements of the third body, where $a$ is the semimajor axis, $e$ the eccentricity,
$i$ the orbital inclination, $\omega$ the argument of perihelion, $\Omega$ the
longitude of the ascending node, $M$ the mean anomaly.
We denote by $a_1$, $e_1$, $i_1$, $\omega_1$, $\Omega_1$,
$M_1$ the Keplerian elements of the secondary body.
In this work we consider a 1:1 mean motion resonance, which occurs whenever
the mean motions of the third and secondary bodies are equal (equivalently,
the periods of revolution are equal).

We first introduce the resonant angles in terms of
the conservative set-up ($\beta=0$). Precisely,
close to the 1:1 MMR, the resonant angle, say $p$, is defined as the difference
of the mean orbital longitudes of the third and secondary bodies:
\beqno
p=\lambda-\lambda_1 \ ,
\eeqno
where $\lambda=M+\tilde \omega$ with $\tilde \omega = \omega+\Omega$,
and similarly for $\lambda_1$. \red{We denote by $a_c$ the value of the semi-major axis of the small body at the 1:1 MMR.
In terms of the semi-major axis of the secondary, the value of $a_c$ within the conservative framework turns out to be
$$
a_c=a_1\ .
$$}
We also remark that the term which corresponds to the solar radiation pressure (i.e.,
\equ{eq:F2} with $\gamma=0$) just contributes to modify the mass parameter of the
central body, \(\mu_0\), by the factor \((1-\beta)\). This implies that we have an
apparent central mass \(\mu_0 (1-\beta)\), instead of the mass \(\mu_0\) in the
equations of motion for the third body. Therefore, for $\beta\neq0$
the resonant value of the semi-major axis, in case of a MMR, is shifted
according to the following relation:
\beqno
a_{res}=(1-\beta)^{1/3}a_c \ ,
\eeqno
where $a_c$ is the nominal value of the semi-major axis of the conservative
case with $\beta=0$.\\

Let us denote by $(L,G,H,\ell,g,h)$ the action--angle Delaunay's variables, which are related to
the orbital elements by (remind that $\mu_*=\mu_0+\mu_1=1$)
\beqano
L&=&\sqrt{a}\ ,\qquad\ \ G=\sqrt{a(1-e^2)}\ ,\qquad\ H=\sqrt{a(1-e^2)}\ \cos i\ ,\nonumber\\
\ell&=&M\ ,\qquad\ \ \ g=\omega\ ,\qquad\qquad\qquad\ \ \ \ h=\Omega\ .
\eeqano
Setting $s=\sin{i\over 2}$, we have that the elements describing the orbit can be expressed in terms
of the Delaunay variables as
\beq{aes}
a={L^2}\ ,\qquad e=\sqrt{1-{G^2\over L^2}}\ ,\qquad s={1\over \sqrt{2}}\ \sqrt{1-{H\over G}}\ .
\eeq
The Hamiltonian function associated to the restricted three--body problem, and expressed in Delaunay variables,
is given by
\beq{ham}
{\mathcal H}(L,G,H,\ell,g,h,\ell_1)=-{(1-\beta\mu_0)^2\over {2L^2}}-\mu_1\
{\mathcal R}(L,G,H,\ell,g,h,\ell_1)\ ,
\eeq
where $\ell_1$ denotes the mean anomaly of the perturber, $\mu_1$ is the mass--ratio of the primaries and
the perturbing function ${\mathcal R}$ is given by the following expression (compare with \equ{UBS}):
\beq{PFrr1psi}
{\mathcal R}={1\over {(r^2+r_1^2-2rr_1\cos\psi)^{1\over 2}}}-{{r\cos\psi}
\over {r_1^2}}-{1\over r}\ ,
\eeq
where $\psi$ is the angle between $(r,r_1)$. We expand $\mathcal R$ around
\red{$\rho=r/r_1-1$} that gives to low orders (see \cite{Lho14}, Appendix B for higher order
expansions):
\beqa{PF11}
{\mathcal R}&\simeq&
\red{-\frac{1}{r}}
-\frac{21 r^2 \cos ^2(\psi )}{64 \sqrt{2} r_1^3}-
\frac{3 r^2 \cos (\psi )}{16\sqrt{2} r_1^3}+
\frac{r^2}{8 \sqrt{2} r_1^3}+
\frac{15 r \cos ^2(\psi )}{32\sqrt{2} r_1^2}\\
&+&\frac{15 \cos ^2(\psi )}{64 \sqrt{2} r_1}+
\frac{r \cos(\psi )}{8 \sqrt{2} r_1^2}-
\frac{r \cos (\psi )}{r_1^2}+
\frac{9 \cos (\psi)}{16 \sqrt{2} r_1}-
\frac{3 r}{4 \sqrt{2} r_1^2}
+ \red{O\left(\rho^3,\cos^3(\psi)\right)} \nonumber \ .
\eeqa
In the next step, the quantities $r$, $r_1$, $\psi$ must be expressed in
terms of the Delaunay variables by standard Keplerian relations
(see, e.g., \cite{alebook, chrbook}). As a consequence, the perturbing function
${\mathcal R}$ can be suitably expanded in Fourier--Taylor series, as
shown in Appendix B.
We are now in the position to introduce the \sl resonant variables \rm
$(P,Q,\red{W},p,q,\red{w})$ as
\beqa{PQR}
P&=&L\qquad\qquad\qquad \ p=\ell-\ell_1+g-g_1+h-h_1\nonumber\\
Q&=&G-L\qquad\qquad \ q=g-g_1+h-h_1\nonumber\\
\red{W}&=&H-G\qquad\qquad \red{w}=h-h_1
\eeqa
with inverse transformation
\beqa{PQRinv}
L&=&P\qquad\qquad\qquad\qquad \ell=p-q+\ell_1\nonumber\\
G&=&P+Q\qquad\qquad\qquad g=q-\red{w}+g_1\nonumber\\
H&=&P+Q+\red{W}\qquad\qquad h=\red{w}+h_1\ .
\eeqa
 Let us denote by \red{$\overline {\mathcal R}$} the average of ${\mathcal R}$
over the mean \red{anomaly} $\ell_1$. Then, the averaged equations in
terms of the resonant variables are given by
\beqa{PQRave}
\dot P&=&\mu_1{{\partial \red{\overline {\mathcal R}}}\over{\partial p}}\ \qquad\qquad
\dot p={{(1-\beta\mu_0)^2}\over P^3}-1-\mu_1{{\partial \red{\overline {\mathcal R}}}
\over{\partial P}}\nonumber\\
\dot Q&=&\mu_1{{\partial \red{\overline {\mathcal R}}}\over{\partial q}}\ \qquad\qquad
\dot q=-\mu_1{{\partial \red{\overline {\mathcal R}}}\over{\partial Q}}\nonumber\\
\dot{\red{W}}&=&\mu_1{{\partial \red{\overline {\mathcal R}}}\over{\partial{\red{w}}}}\ \qquad\qquad
\dot{\red{w}}=-\mu_1{{\partial \red{\overline {\mathcal R}}}\over{\partial{\red{W}}}}\ .
\eeqa
\red{The term -1 in $\dot p$ stems from the fact that the transformation \equ{PQR}
depends
on time through $\ell_1$, appearing in the definition of $p$.}
Equations \equ{PQRave} represent the contribution of the conservative part to
which we must add the effect of the dissipation due to the Poynting--Robertson
drag.
Precisely, we add the dissipation averaged over the mean anomaly. Since the
average dissipation is zero for the angle variables (see \cite{JL}), the
Poynting--Robertson effect contributes to the equations \equ{PQRave} only by
modifying the equations of the action variables according to the following
formulae:
\beqa{disseq}
\dot P&=&\mu_1{{\partial \red{\overline {\mathcal R}}}\over{\partial p}}+Y_P\ \qquad\qquad
\dot p={{(1-\beta\mu_0)^2}\over P^3}-1-\mu_1{{\partial \red{\overline {\mathcal R}}}
\over{\partial P}}\nonumber\\
\dot Q&=&\mu_1{{\partial \red{\overline {\mathcal R}}}\over{\partial q}}+Y_Q\ \qquad\qquad
\dot q=-\mu_1{{\partial \red{\overline {\mathcal R}}}\over{\partial Q}}\nonumber\\
\dot{\red{W}}&=&\mu_1{{\partial \red{\overline {\mathcal R}}}\over{\partial{\red{w}}}}+\red{Y_W}\ \qquad\qquad
\dot{\red{w}}=-\mu_1{{\partial \red{\overline {\mathcal R}}}\over{\partial{\red{W}}}}\ ,
\eeqa
where
\beqano
Y_P&=&Y_L\nonumber\\
Y_Q&=&Y_G-Y_L\nonumber\\
\red{Y_W}&=&Y_H-Y_G
\eeqano
and $Y_L$, $Y_G$, $Y_H$ are defined as follows.
Denoting by $n$ the mean motion of the third body,
from \cite{JL} we have the following expressions:
\beqa{YLHG}
Y_L&=&-\mu_0\beta n{{1+{3\over 2}e^2}\over {c(1-e^2)^{3\over 2}}}\nonumber\\
Y_G&=&-{{\mu_0\beta n}\over c}\nonumber\\
Y_H&=&-{{\mu_0\beta n}\over c}\cos i\ .\nonumber\\
\eeqa
The effect of the dissipation on the orbital elements can be evaluated using the expressions \equ{aes},
computing the time derivative of the orbital elements and inserting \equ{disseq} and \equ{YLHG} in place of $\dot L$,
$\dot G$, $\dot H$. More precisely, we obtain:
\beqa{aei}
{{da}\over {dt}}&=& {{2L\dot L}}=-{{\sqrt{a}(1+3e^2)\mu_0\beta n}\over {c(1-e^2)^{3\over 2}}}\nonumber\\
{{de}\over {dt}}&=& {G\over {L^2 e}}({G\over L}\dot L-\dot G)={{\sqrt{1-e^2}\ \mu_0\beta n}\over {\sqrt{a}\ e\ c}}
-{{(2+3e)\ \mu_0\beta n}\over {2\sqrt{a}\ e\ c(1-e^2)^{1\over 2}}}\nonumber\\
{{di}\over {dt}}&=& -{1\over {G^2\sqrt{1-{H^2\over G^2}}}}(\dot H G-H\dot G)=0
\eeqa
(the last result comes from the fact that $Y_HG-HY_G=0$, being $H/G=\cos i$).

The above equations show that the sole dissipation drives to circular orbits
(i.e., $e=0$), which end up to collide with the primary body (i.e., $a=0$), while
no effect is performed on the inclination.

\begin{remark}
To evaluate the occurrence of stationary solutions, we can make
use of Tisserand criterion (\cite{Moulton}, notice that the computation is valid for internal,
1:1 or external resonances). Precisely, we start by mentioning that under the solar radiation pressure
the Jacobi constant is given by
$$
C={{(1-\beta\mu_0)}\over a}+2\sqrt{(1-\beta\mu_0)a(1-e^2)}\ \cos i\ .
$$
Recalling the last result in \equ{aei}, we have that
$$
{{dC}\over {dt}}=-{{1-\beta\mu_0}\over a^2}{{da}\over {dt}}+{{(1-\beta\mu_0)\cos i}\over {2\sqrt{(1-\beta\mu_0)a(1-e^2)}}}\
((1-e^2){{da}\over {dt}}-2ae{{de}\over {dt}})\ .
$$
Using the first two expressions in \equ{aei}, we obtain
\beqano
{{dC}\over {dt}}&=&{{\mu_0\beta n}\over {c\ a^{3\over 2}\ (1-e^2)^{3}}}\Big[
-2 a^{3\over 2}(1-e^2)^3\sqrt{1-\mu_0\beta}\cos i+(1-e^2)^{3\over 2}(2+3e^2)(1-\mu_0\beta)\Big]\nonumber\\
&=&{{\mu_0\beta n\sqrt{1-\mu_0\beta}}\over {c\ a^{3\over 2}\ (1-e^2)^{3\over 2}}}\Big[
-2 a^{3\over 2}(1-e^2)^{3\over 2}\ \cos i+(2+3e^2)\sqrt{1-\mu_0\beta}\Big]\ .
\eeqano
The condition that the Jacobi integral is constant, i.e. $dC/dt=0$, implies in the limit $e=0$ that
\beq{C}
a^{3\over 2}\ \cos i=\sqrt{1-\mu_0\beta}\ .
\eeq
Given that Kepler's third law under solar radiation pressure reads as
\beq{na}
n^2a^3=(1-\beta\mu_0)^4\ ,
\eeq
in a 1:1 MMR (i.e., with $n=1$) we have that \equ{C} reduces to
\beq{ci}
\cos i={{n}\over {(1-\mu_0\beta)^{3\over 2}}}\ ,
\eeq
which can be satisfied only if $n\leq(1-\mu_0\beta)^{3\over 2}$.
As it is well known (\cite{Bea94}), this implies that stationary solutions in a 1:1 MMR
with non-zero eccentricity and inclination can only exhibit temporary trapping.

When $\beta=0$, from \equ{ci} we obtain that the Jacobi integral is preserved just for $n\leq 1$,
which corresponds to the small particle on an orbit external to that of the secondary (compare with \cite{Bea94}).
When $\beta\not=0$, this condition is modified and only some external orbits can be considered, precisely those
satisfying $n\leq(1-\mu_0\beta)^{3\over 2}$.
\end{remark}

\section{Stationary solutions}
\label{StaSol}

In order to find stationary solutions of the averaged problem, we look for the
equilibrium solutions associated to \equ{disseq}. More precisely, we fix a set
of parameters $(\mu_1,e_1,s_1,\beta)$, where $s_1=\sin{i_1\over 2}$ with $i_1$
denoting the inclination of the secondary. We determine a set of initial
conditions $(P_0,Q_0, \red{W_0},p_0,q_0,\red{w_0})$, such that the right hand sides of
\equ{disseq} are identically zero for the selected parameter values. We then
back-transform them into the stationary orbital elements
$a_*,e_*,i_*,p_*,q_*,\red{w_*}$.\footnote{\red{To test our numerical approach we also
derive first order formulae in $\beta$ for $a_*$, $p_*$, $e_*$, $q_*$ in
the following way: first, we substitute the ansatz $a_*=a_{res}+C_1\beta$,
$p_*=\pm60^o+C_2\beta$ into ($\dot P$, $\dot p$) of \equ{disseq} to obtain $C_1$,
$C_2$ by setting $e=q=i=w=0$. Next, we use the ansatz $e_*=e_1+C_3\beta$ and
$q_*=\pm60^o+C_4\beta$ to obtain $C_3$ and $C_4$ from ($\dot Q$, $\dot q$)
of \equ{disseq} using the solutions for $a_*$ and $p_*$ we obtained before,
and setting $i=w=0$. No perturbative approach has been used to find $i_*$
and $w_*$. The expansions are shown on top of the respective
figures.}} \red{We notice that in the conservative setting the
equilibria $L_4$ and $L_5$ are mirror symmetric with respect to $\vec r_1$.
Thus, for $\beta=0$ the equilibrium $L_4$ that is given by ($a_*$, $e_*$,
$i_*$, $+p_*$, $+q_*$, $+w_*$) maps into the equilibrium $L_5$ in terms of
($a_*$, $e_*$, $i_*$, $-p_*$, $-q_*$, $-w_*$). Since the derivatives of the
perturbing function with respect to the resonant angles introduce the sine
function into the right hand sides of \equ{disseq} for $\dot P$, $\dot Q$,
$\dot W$, then a small deviation from $L_4$ is symmetrically mapped into a small
deviation from $L_5$. This provokes that the right hand sides of $\dot P$, $\dot Q$,
$\dot W$ in \equ{disseq} have opposite signs~\footnote{\red{In contrast, the
evaluations of the right hand sides of $\dot p$, $\dot q$,  $\dot w$ will
result in terms with same signs for $L_4$ and $L_5$.}} with respect to the
right hand sides evaluations close to $L_4$ and viceversa. However,  in the dissipative case,
mapping small deviations from $L_4$ symmetrically into the vicinity of $L_5$
does not alter the signs of the dissipative terms \equ{YLHG}. Therefore, since
for $\beta\neq0$ the sum of the conservative and dissipative terms must cancel out
to fulfill the requirement for the equilibrium  $\dot P = \dot Q = \dot W = 0$, then
the respective terms will not balance themselves in the same way close to $L_4$
in comparison to $L_5$. Henceforth, we can expect an asymmetry of the equilibria
$L_4$ and $L_5$ in presence of PR-drag.} To evaluate the context of the
different dimensions of the $6$-dimensional phase space, we perform the
calculations in the planar and spatial versions of the circular and elliptic
restricted three-body problems.  We start with the circular-planar case of
Section~\ref{sec:CPRTBP}, then we let the orbits be inclined as in
Section~\ref{sec:SCRTBP}, we analyze the elliptic-planar case in
Section~\ref{sec:ERTBP} and finally we discuss in Section~\ref{sec:SERTBP} the
most general model.  The different settings are referred to by appropriate
acronyms given at the beginning of each section.

\subsection{Circular-planar case (CPRTBP)}\label{sec:CPRTBP}
\begin{figure}[h]
\centering
\includegraphics[height=.43\linewidth]{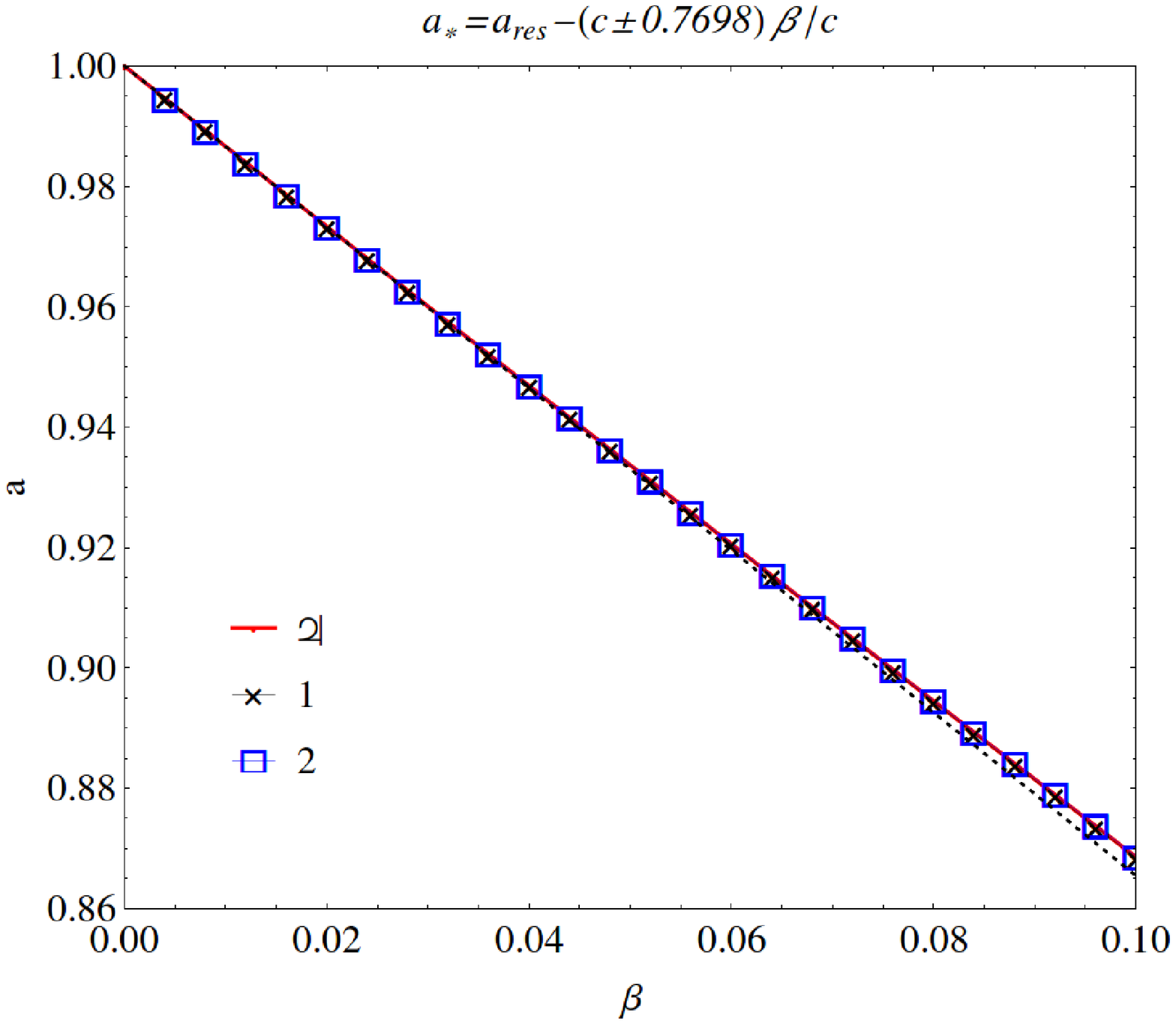}
\includegraphics[height=.43\linewidth]{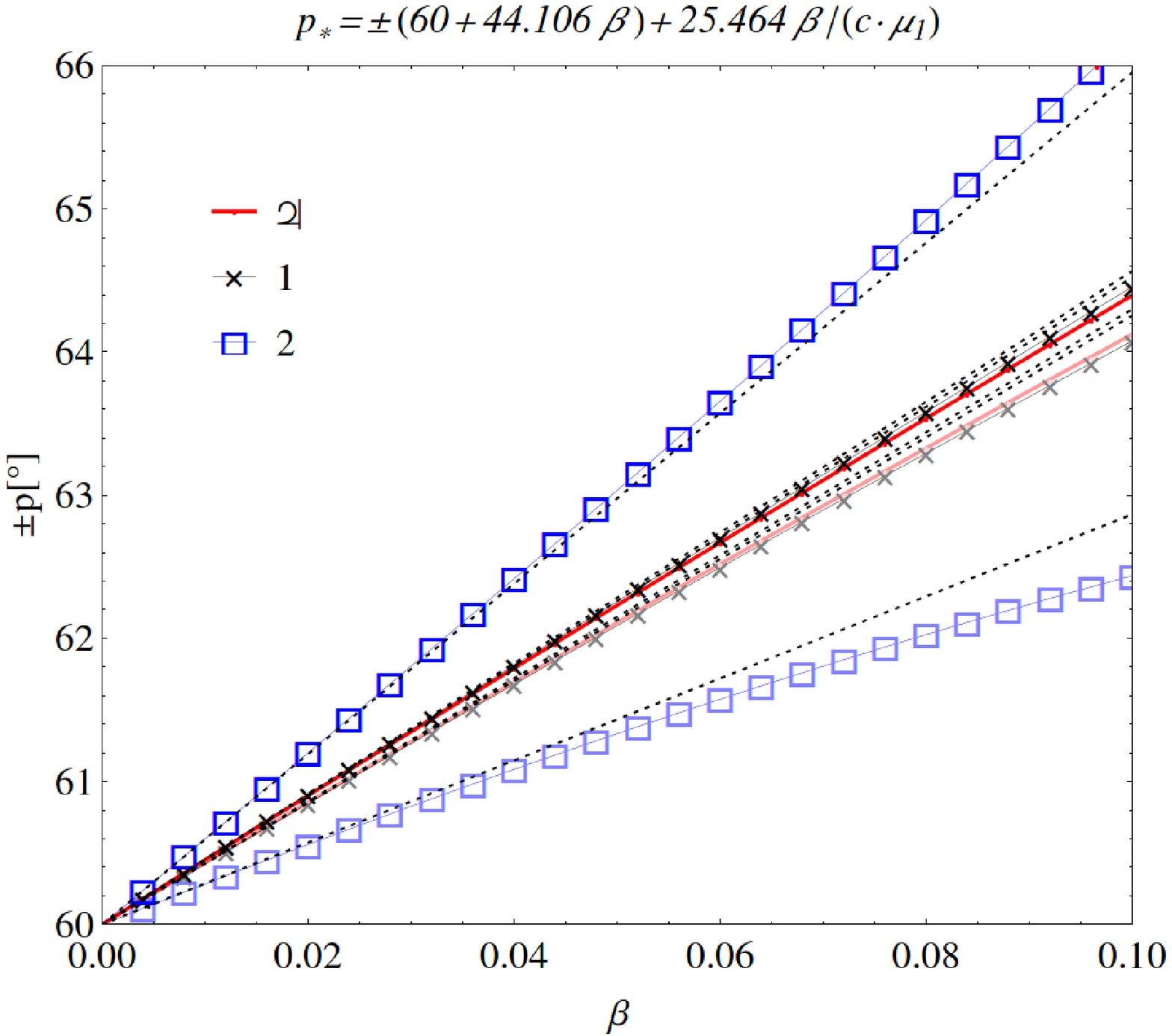}
\caption{Variation of the equilibrium solutions \red{for $L_4$ (dark) and $L_5$
(light)} in the CPRTBP for $a$ (left) and  $p$ (right) as a function of the
parameter $\beta$ for different mass parameters $\mu_1/\mu_J$ equal to $1$ (red
dot), $1.5$ (black cross), $0.6$ (blue square), respectively. \red{On the left,
$L_4$ and $L_5$ overlap.} \red{Dotted lines correspond to first order formulae.}}
\label{fig:crtbp}
\end{figure}

We assume that the secondary moves on a circular orbit, while the third body
may have non-zero eccentricity, and that all bodies move on the same plane.
Therefore we set $e_1,i_1,i,\Omega=0$ in $\mathcal R$ to obtain $\red{\overline{\mathcal R}}$,
and we immediately find $\red{\dot W}=0, \red{\dot w}=0$ in the equations of motion \equ{disseq}. We remark
that
\beq{eq:YQ}
Y_Q=Y_G-Y_L=-{{\mu_0\beta n}\over {c}}\Big(1-{{1+{3\over 2}e^2}
\over {(1-e^2)^{3\over 2}}}\Big)\ .
\eeq
This implies that $Y_Q=0$ whenever $\beta=0$ (the usual conservative case) or if $e=0$;
thus, in the dissipative setting, if we set $e=0$, we are reduced to find the
solution just of the system of equations
\beq{Pp}
\dot P=0\ , \qquad
\dot p=0\ ,
\eeq
since $\dot Q=0$ also in the conservative case, and we cannot solve for $q_*$, since
the angle $q$ is an ignorable variable also in the dissipative case
due to the fact that \red{$e=0$}. We thus neglect the equation $\dot q=0$.\footnote{We remark, that in presence of
dissipation, if $e\not=0$ we have $\dot Q\not=0$ for $\beta\not=0$ and thus
$\dot e\not=0$, while in the conservative set-up ($\beta=0$) we find $\dot e=0$
and the eccentricity is a conserved quantity. Therefore, $e$ is not a conserved
quantity anymore in presence of PR drag in the circular problem.} The system
of equations \equ{Pp} only provides the equilibrium solution for the
variables $P$ and $p$; in particular, the solution for $P$ gives the
equilibrium value of the semi-major axis. In Figure~\ref{fig:crtbp} we report
the variation of the equilibrium solutions for $a$ and $p$, for different values of
$\mu_1$, as a function of the parameter $\beta$, which varies in the interval
$[0,0.1]$.

From Figure~\ref{fig:crtbp} we see, that for $\beta=0$ we recover the equilibrium
solution of the conservative case at $a_*=1$, $p_*=60^{\circ}$. For $\beta\not=0$
the equilibrium $a_*$ decreases \red{below} $0.88$ for all different mass parameters
$\mu_1$, while the value of the resonant argument $p_*$ strongly depends on
the choice of \red{$c\cdot\mu_1$} (ranging from about $64^{\circ}$ for $\beta=0.1$ and
$\mu_1=\mu_{J}$ to $66^{\circ}$ for $\beta=0.1$ and $\mu_1=0.6\mu_J$ \red{for
$L_4$ and $-62^{\circ}$ to $-63.5^{\circ}$ for $L_5$}).

\subsection{Spatial-circular case (SCRTBP)}\label{sec:SCRTBP}
In this model we assume that the secondary moves on a circular orbit, the third
body may have non-zero eccentricity, and that both smaller bodies move on
inclined planes. We thus set $e_1,\omega_1=0$, and keep $i,\Omega\not=0$ in
$\mathcal R$ to obtain $\red{\overline{\mathcal R}}$.  Like in the CPRTBP (see
\equ{eq:YQ}), we find that the equation for $\dot Q$ can only be solved for
$\beta=0$ or $e=0$ also in the spatial case. However, contrary to the CPRTBP,
we have $\dot {\red W}\not=0$, $\dot {\red w}\not=0$ in the system of equations \equ{disseq},
and we are led to solve the equations of motion for the variables $P$, $p$,
${\red W}$, ${\red w}$, simultaneously. For $a_*, p_*$ we find the same equilibrium values as
in the CPRTBP. For $\beta=0$ we recover the known conservative equilibrium
value of the Lagrange orbit\footnote{In this case the equilibrium solutions are
replaced by periodic orbits and consequently we speak more appropriately of a
\sl Lagrange orbit. \rm} at $i_*=i_1$, and ${\red w}_*=0$. The same equilibrium
positions are found for $\beta\not=0$: the values for the spatial variables
correspond to the case where the two bodies share their lines of nodes, while
the relative inclination turns out to be zero. We conclude that the Lagrange
orbits in the SCRTBP can be identified with the Lagrange orbits in the CPRTBP
to which can be related by simple rotations. However, in the dissipative case
we find additional equilibrium solutions, such that the equilibrium inclination
$i_*$ is different from that of the secondary $i_1$ with large ${\red w}_*\not=0$.
Since we focus our study on the Lagrange configuration (with ${\red w}_*\simeq0$), we did
not investigate them further. \red{We also remark that an additional class of
equilibria can be artificially constructed in the following way. We premise that the
solution of the system of equations $\dot P=\dot Q=\dot {\red W}=\dot p=\dot q=\dot {\red w}=0$
for $i\not= i_1$, leads to possible equilibria with very large ${\red w}$, which is not
consistent with the expected physical picture. Instead of solving for all variables,
we can fix ${\red w}=0$ in \equ{disseq} and solve for the reduced system $\dot P=\dot Q=\dot p=\dot q=0$,
thus leading to an equilibrium solution $a_*$, $e_*$, $p_*$, $q_*$ for $i\not= i_1$.
It turns out that for a moderate difference of the inclination from $i_1$ and for small values of $\beta$,
the value of $a_*$, $e_*$ are not much altered, while a bigger difference is found for $p_*$, $q_*$,
when compared to the case $i=i_1$.} \\

When we plot the graphs of the equilibrium solutions for $i_*$,
${\red w}_*$, starting for example with $i_1=5^o$, as a function of the
parameter $\beta$, varying in the interval $[0,0.1]$ for different
mass ratios $\mu_1$, we notice that $i_*=i_1$, ${\red w}_*=0$ holds true
for arbitrary $\beta$. The plots for $a_*$, $p_*$ overlap to those
of Figure~\ref{fig:crtbp}. \\


\subsection{Elliptic-planar case (EPRTBP)}\label{sec:ERTBP}
We assume that the eccentricity of the smaller primary is different from zero,
but we make again the assumption that all bodies move on the same plane, like we
already did in the CPRTBP. We thus set $i,i_1,\Omega,\Omega_1=0$, but we keep
$e,e_1,g,g_1\not=0$ in $\mathcal R$ to obtain $\red{\overline{\mathcal R}}$. Like in
the CPRTBP we find $(\dot {\red W}=0, \dot {\red w}=0)$, also for non-zero $\beta$ in
the system \equ{disseq}, and we are thus led to solve a system of equations in
the four coordinates $P$, $Q$, $p$, $q$, where we must account for the fact
that the dissipation acts only on the action variables $P$ and $Q$.\\

\begin{figure}[h]
\centering \vglue0.1cm \hglue0.2cm
\includegraphics[height=.42\linewidth]{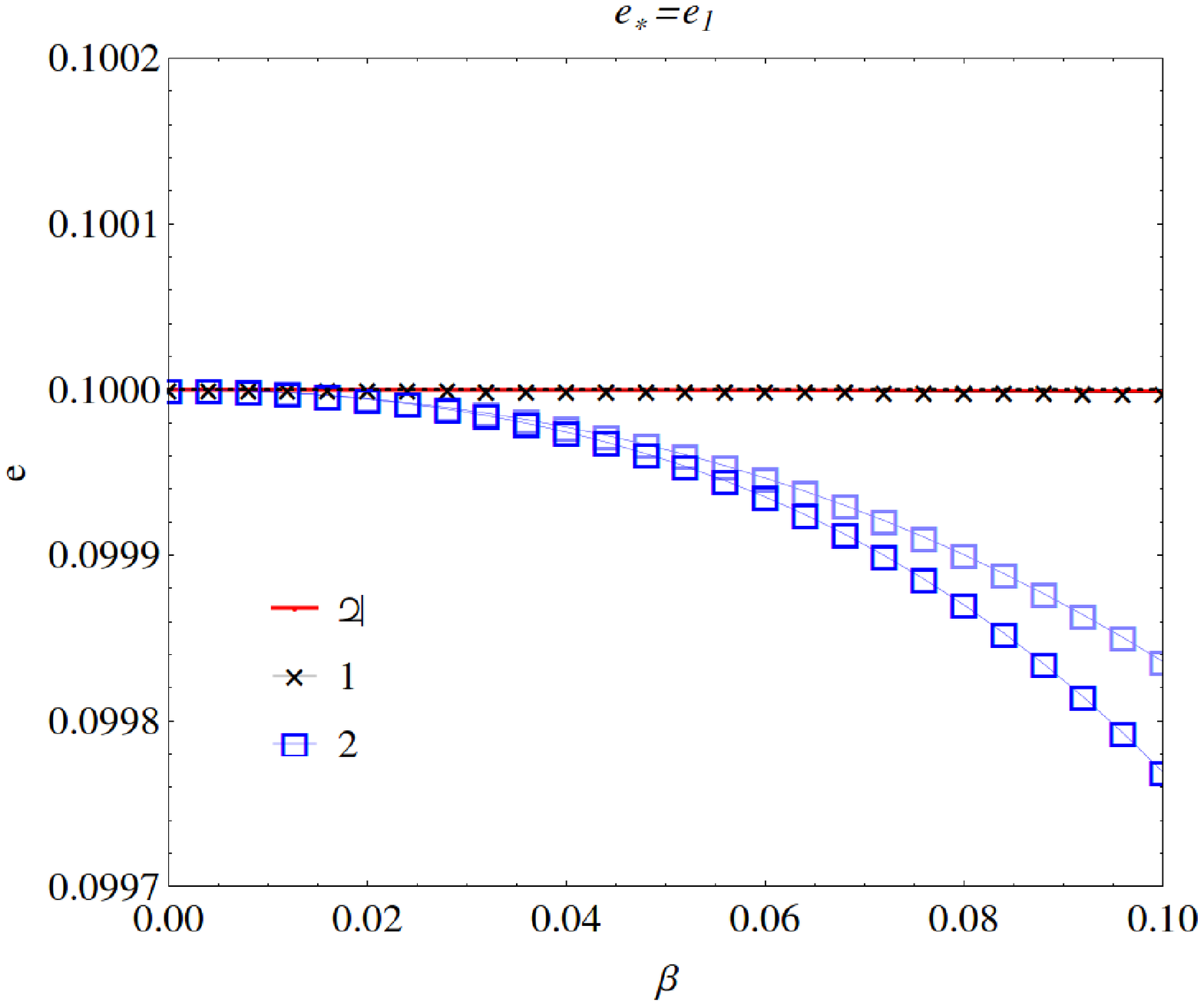}
\includegraphics[height=.42\linewidth]{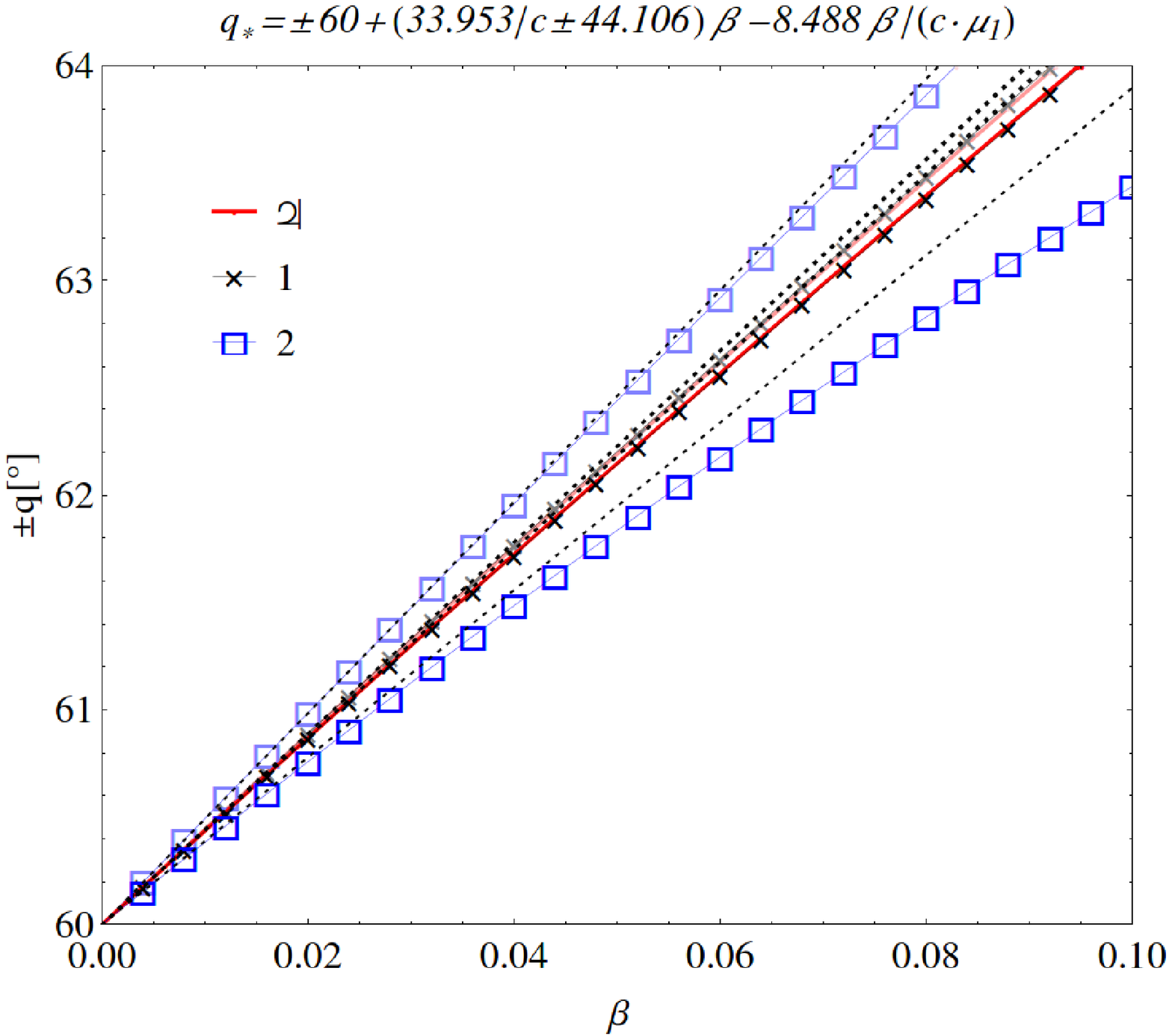}
\caption{Variation of the equilibrium solutions \red{for $L_4$ (dark) and
$L_5$ (light)} in the EPRTBP for
$e$ (left), and $q$ (right) as a function of the parameter $\beta$
for $e_1=0.1$ and different mass ratios $\mu_1/\mu_J$ equal $1$
(red dot), $1.5$ (black cross), and $0.6$ (blue square),
respectively. The plots for $a$, $p$ coincide with those of
Figure~\ref{fig:crtbp}.
\red{On the left $L_4$ and $L_5$ overlap for $\mu_J$ and
$1.5\mu_J$.} \red{Dotted lines correspond to first order formulae.}}
\label{fig:ertbp}
\end{figure}

We find that the correlations between $a_*$, $p_*$ and $\beta$ remain the same
as in the CPRTBP and SCRTBP; we therefore omit the corresponding figures here.
We only report in Figure~\ref{fig:ertbp} the graphs of the equilibrium
solutions for $e_*$ and $q_*$ as a function of the parameter $\beta$. We find
that $e_*$ depends on the parameter $\beta$, while we have even a stronger
dependency of the equilibrium value for the angle $q_*$ with $\beta$. For
$\beta=0$ we have $e_*=e_1$ and $q_*=60^{\circ}$. For large enough $\mu_1$ the
equilibrium $e_*$ remains the same, while $q_*$ tends to $64^{\circ}$ for
$\beta=0.1$. It is interesting to notice that for $\mu_1$ small (blue in
Figure~\ref{fig:ertbp}) $e_*$ tends to smaller values (still close to
$e_*=e_1$), while the effect on $p_*$ is smaller than for larger masses of $\mu_1$.
\red{We also remark, that while in Figure~\ref{fig:crtbp} (right) the solution
for $L_4$ tends to larger values, in Figure~\ref{fig:ertbp} (right) the solution
for $L_5$ tends to larger ones. }

\subsection{Spatial-elliptic case (SERTBP)}\label{sec:SERTBP}
In the most general case, we assume that the secondary moves on an elliptic
orbit and on an inclined plane, so that both $e_1$ and $s_1$ are different from
zero. We thus investigate the full dynamics of the equations \equ{disseq}, where
we keep all orbital elements in $\mathcal R$ to obtain $\red{\overline{\mathcal R}}$.
For $\beta=0$ we find the real
equilibrium solution at $a_*=a_1=1$, $e_*=e_1$, $i_*=i_1$, $p_*=q_*=60^{\circ}$,
and \red{$w_*=0$}, that corresponds to the well-known Lagrange orbit: the orbital
planes share their line of nodes with zero relative inclination, while the line
of apsides of the third body is rotated by $60^{\circ}$, and the difference
in orbital longitudes is $60^{\circ}$.
For $\beta\not=0$ we solve for the system of equations in all variables
$(P,Q,\red{W},p,q,\red{w})$ and we obtain that the equilibrium solution is typically
obtained when $e=e_1$ and $s=s_1$.  Indeed, the equation for $\dot Q$ does not
depend on $s_1$ and it is zero for $e=e_1$. On the other hand, the equation for
\red{$\dot W$} does not depend on $e_1$ and it becomes zero only when $s=s_1$.\\

\begin{figure}[h]
\centering 
\includegraphics[height=.36\linewidth]{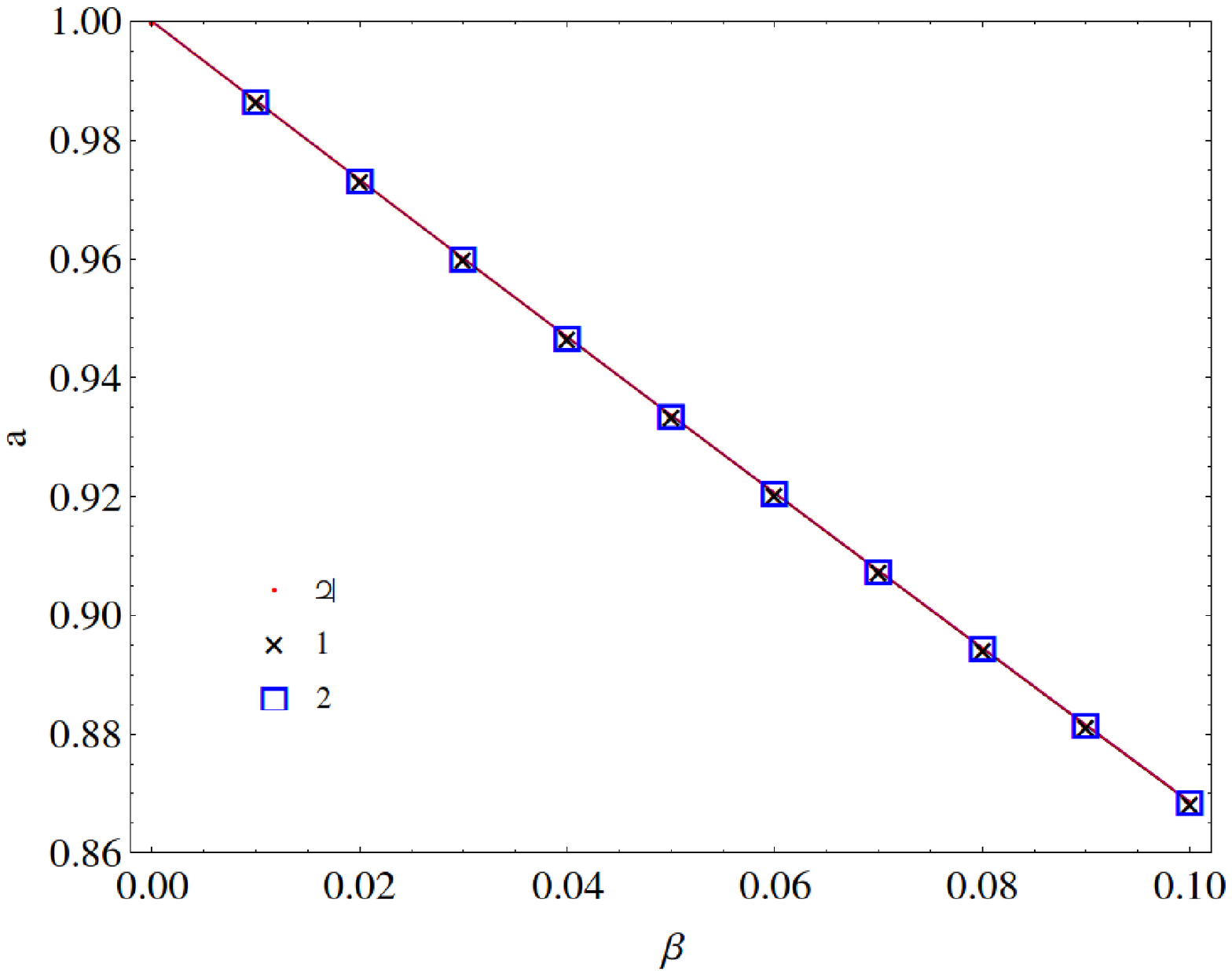}
\includegraphics[height=.36\linewidth]{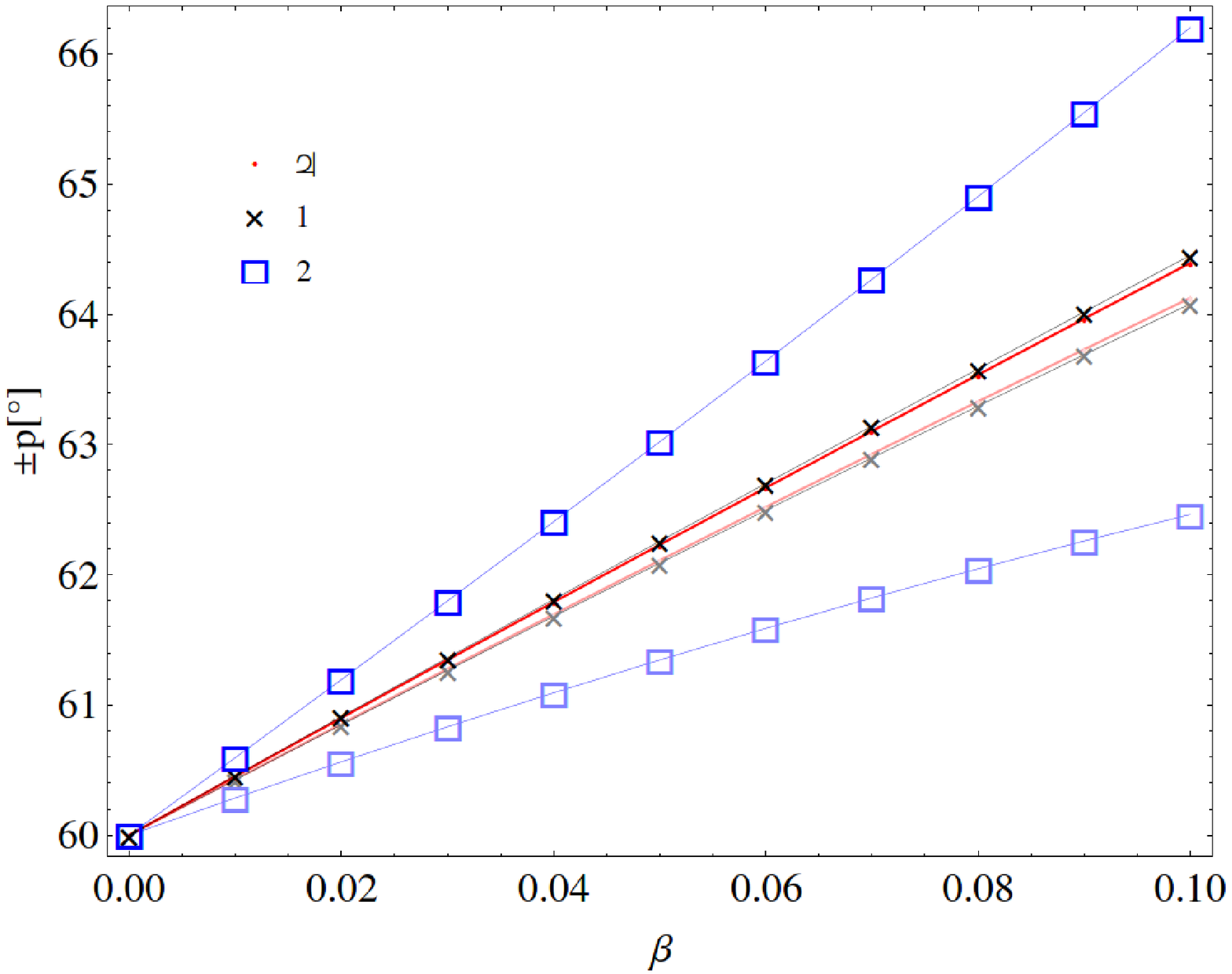} \\
\includegraphics[height=.36\linewidth]{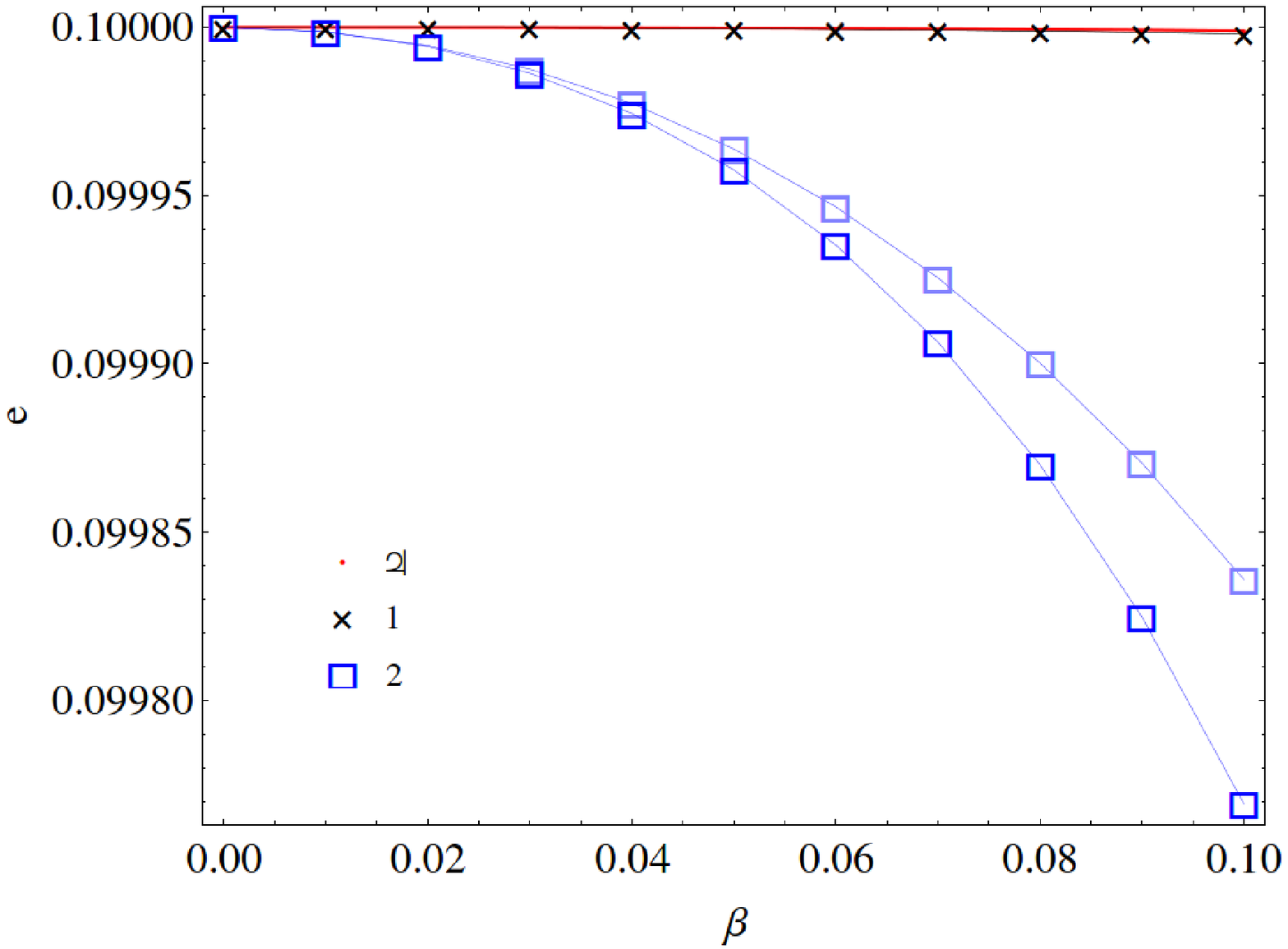}
\includegraphics[height=.36\linewidth]{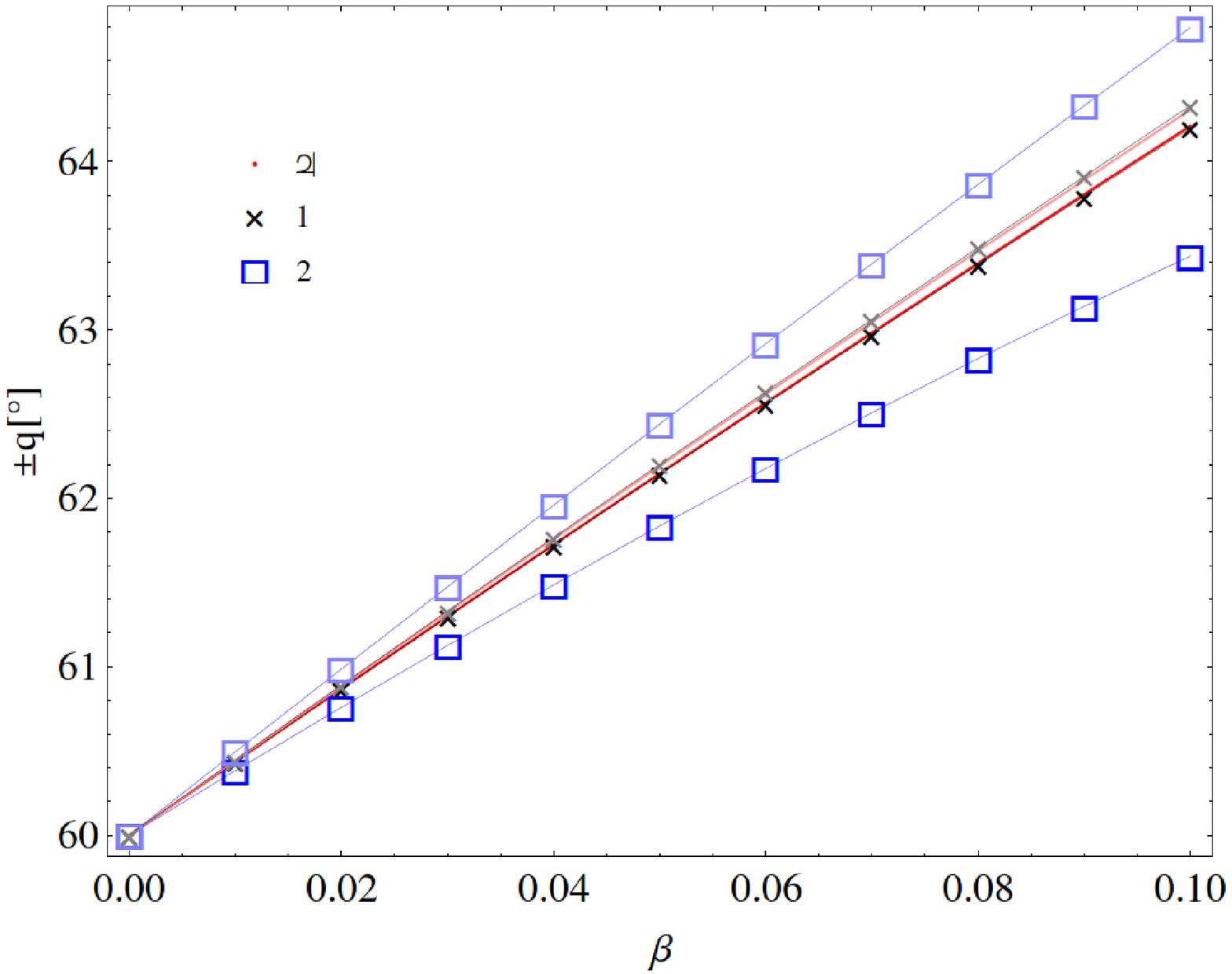} \\
\includegraphics[height=.36\linewidth]{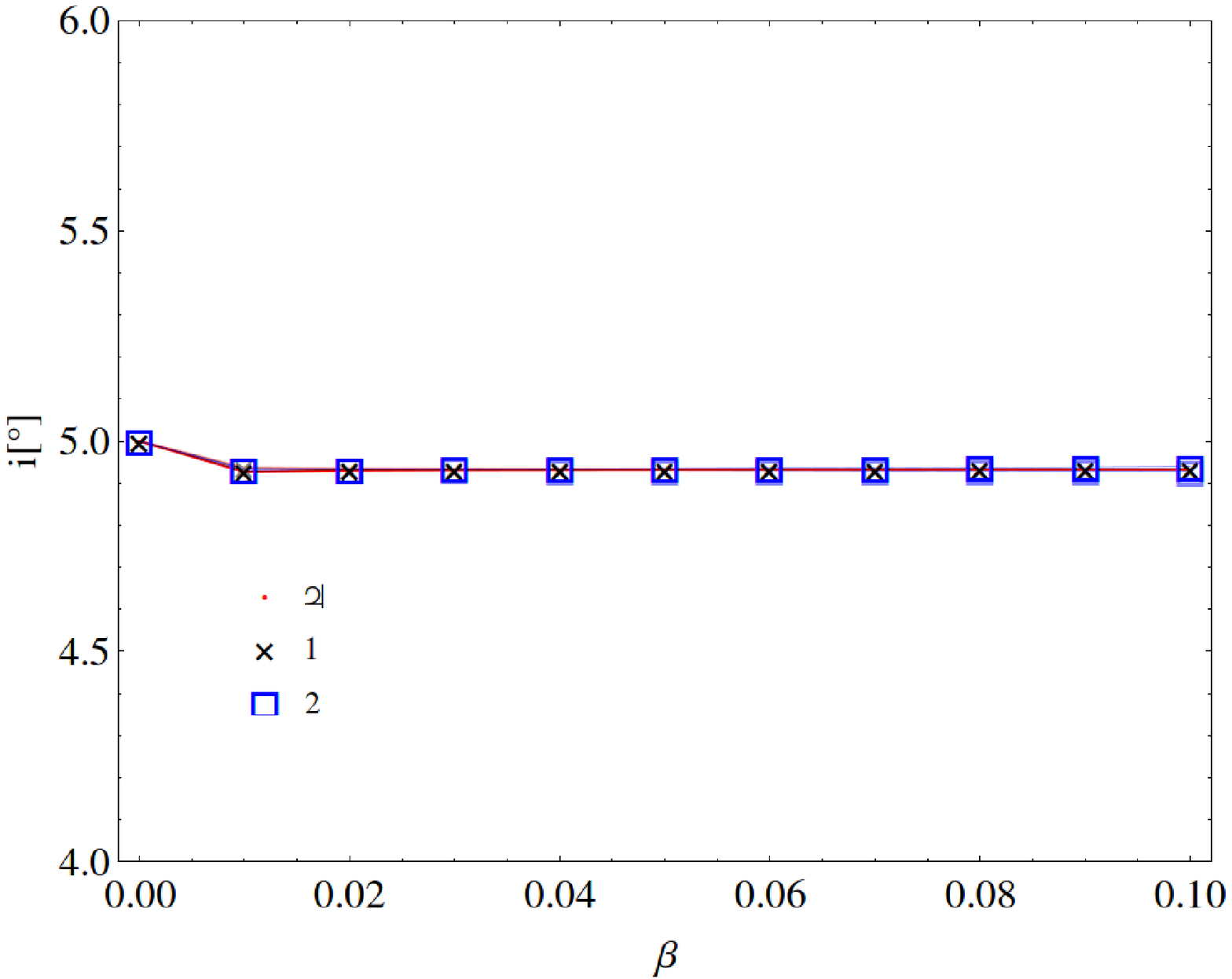}
\includegraphics[height=.36\linewidth]{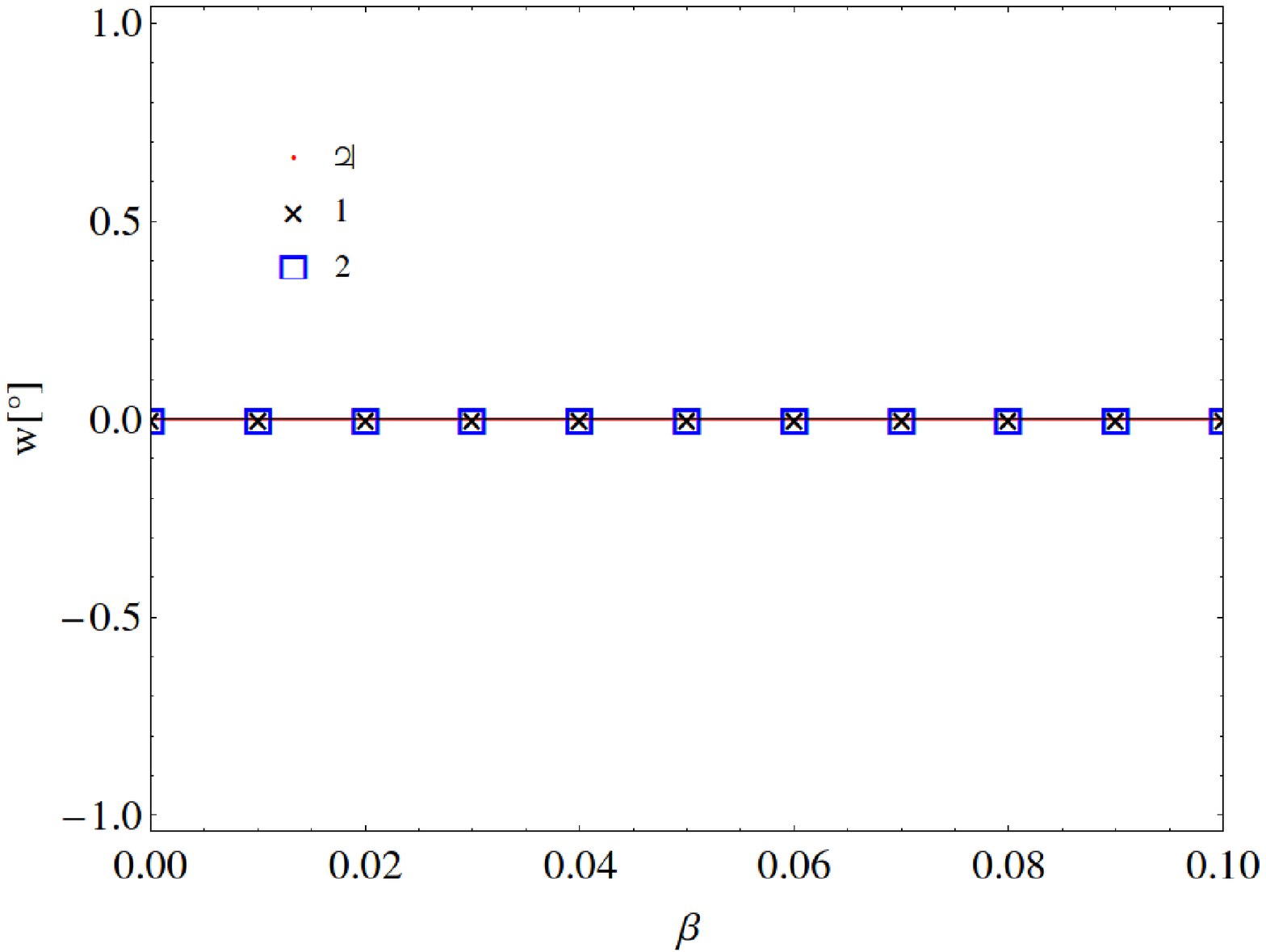}
\caption{Variation of the equilibrium solutions \red{for $L_4$ (dark) and
$L_5$ (light)} in the SERTBP for
$e_1=0.1$, $i_1=5^{\circ}$ and for different mass parameter
$\mu_1/\mu_J$ equal $1$ (red dot), $1.5$ (black cross), $0.6$
(blue square), for $a$ (upper left panel), $p$ (upper right
panel), $e$ (middle left panel), $q$ (middle right panel), $i$
(bottom left panel), \red{$w$} (bottom right panel) as functions of the
parameter $\beta$. \red{Light curves overlap with dark ones if not
visible.}} \label{fig:sertbp}
\end{figure}

We report in Figure~\ref{fig:sertbp} the graphs of the equilibrium solutions
for $a$, $e$, $s$, $p$, $q$  and \red{$w$} for different mass parameters $\mu_1$
as functions of the parameter $\beta$, varying in the interval $[0,0.1]$.

In comparison with Figures~\ref{fig:crtbp}--\ref{fig:ertbp} we find that from a
qualitative point of view the correlations of $a,e,p,q$ with $\beta$ at
their equilibrium values remain the same. We notice, that in the SERTBP,
for $\beta\neq0$ the inclination $i_*$ tends to slightly lower values
than $i_1=5^{\circ}$ for \red{$w_*$} fixed at $0$ degrees. \red{No difference
between $L_4$ and $L_5$ is visible with respect to $i$ and $w$.} \\

\red{We observe in Figures~\ref{fig:crtbp}-\ref{fig:sertbp} that the difference
in the locations of the equilibria in the parameter space between the cases
Jupiter and case 1 is small ($10^{-2\ o}$ for $p$ and $10^{-3\ o}$ for $q$ with
$\beta=0.05$) compared to case 2 ($10^{-1\ o}$ for $p$ and $10^{-2\ o}$ for $q$
with $\beta=0.05$). A possible explanation is as follows: terms in \equ{disseq}
entering proportionally to $\mu_1$ need to be balanced with the dissipative terms that enter with
proportionality factor $\mu_0\beta n/c$. Due to our special choice of units, this term is
proportional to $(1-\mu_1)\beta a_1n_1\, n$. From Table~\ref{t:sys},
with $n_1=2\pi/P_1$, we find that the terms \equ{YLHG} for
Jupiter and case 1 are of the same order of magnitude, while for case 2 the corresponding
term turns out to be one order of magnitude bigger. We can therefore expect that the deviation
of the equilibria from the conservative solution for Jupiter and case 1 are comparable, while
the deviation for case 2 is larger.}

\section{On the behavior of the eigenvalues of the equilibrium positions}
\label{LinSta}

\begin{figure}
\centering
\includegraphics[width=0.46\linewidth]{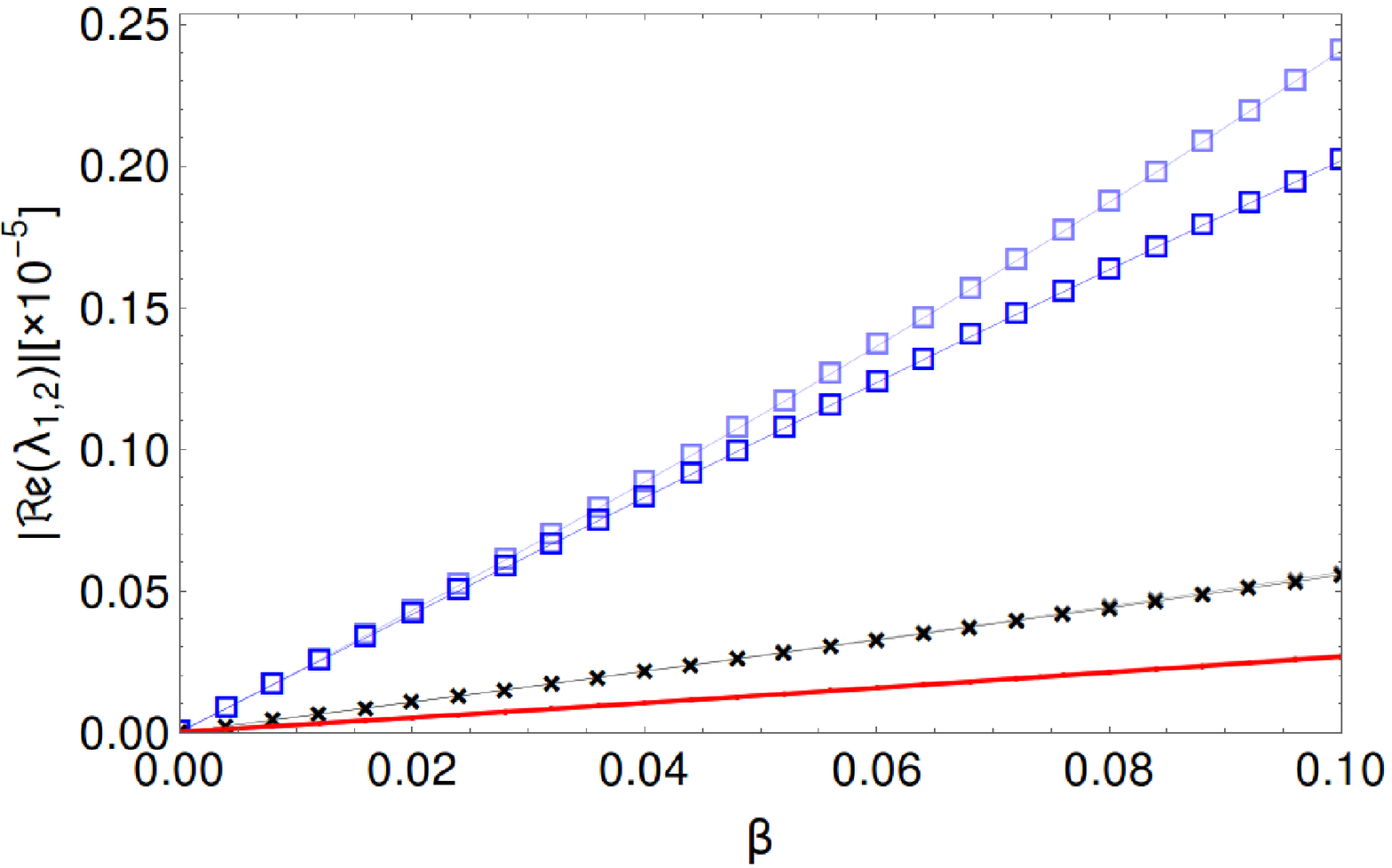}
\includegraphics[width=0.46\linewidth]{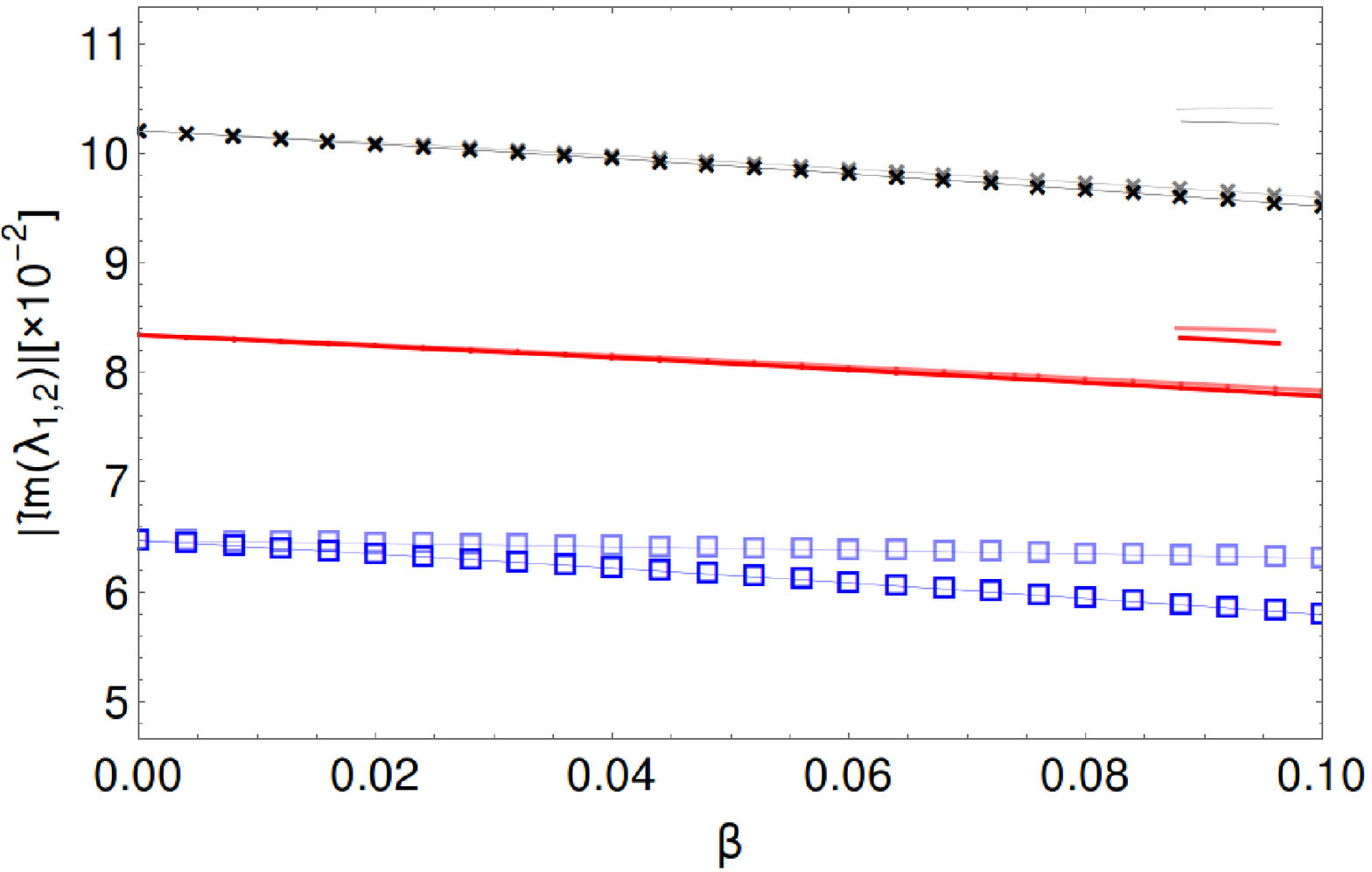} \\
\includegraphics[width=0.46\linewidth]{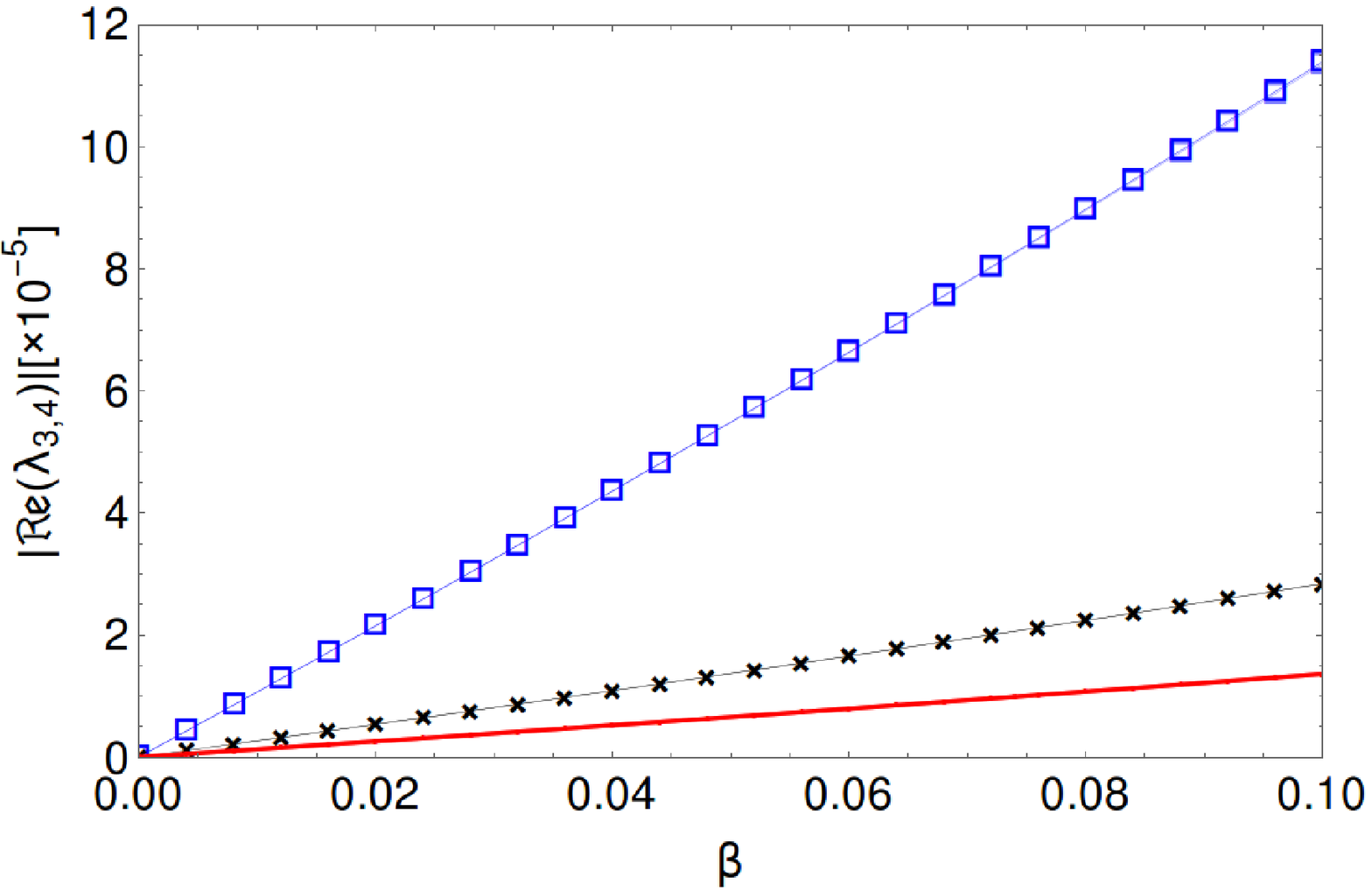}
\includegraphics[width=0.46\linewidth]{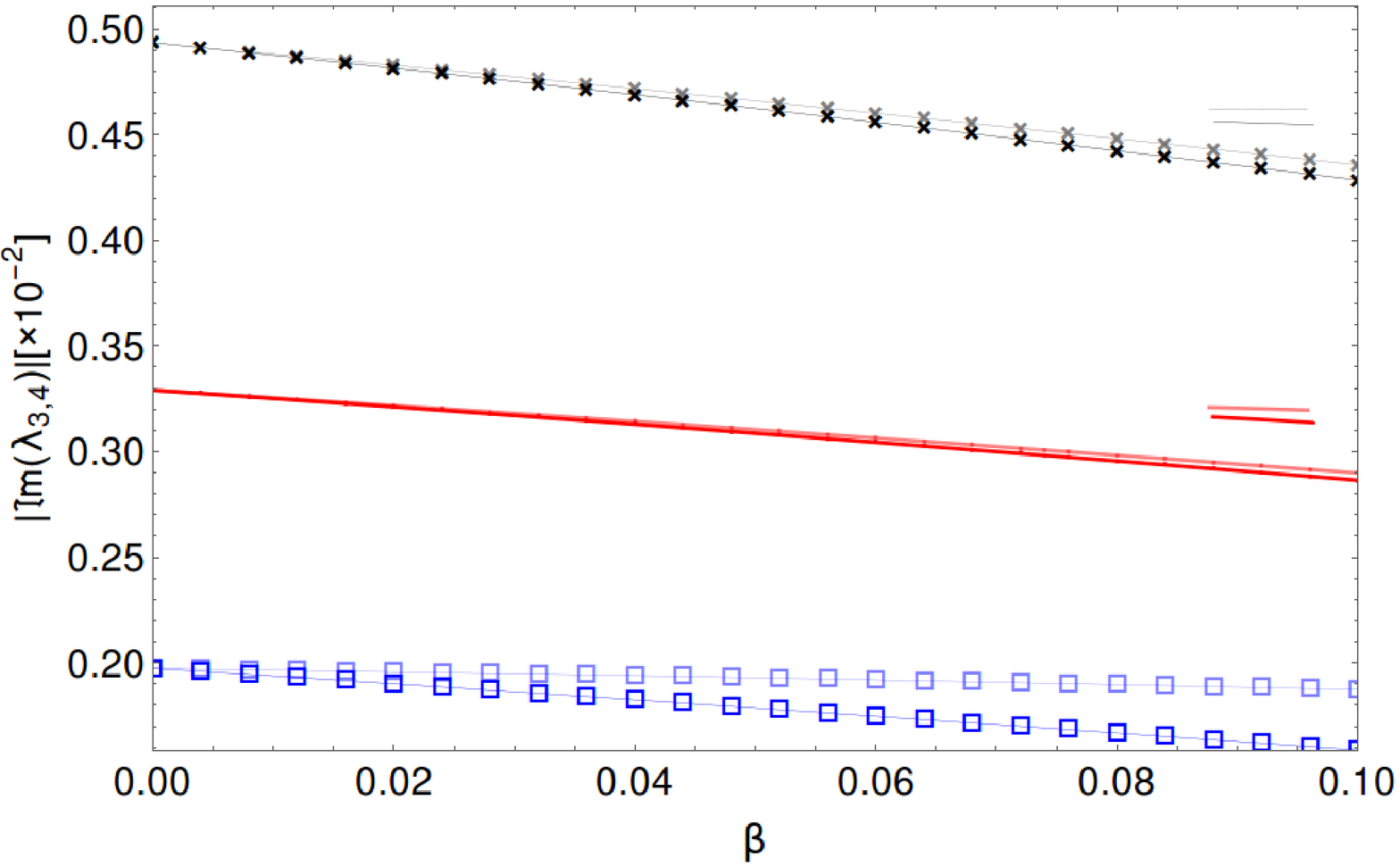} \\
\includegraphics[width=0.46\linewidth]{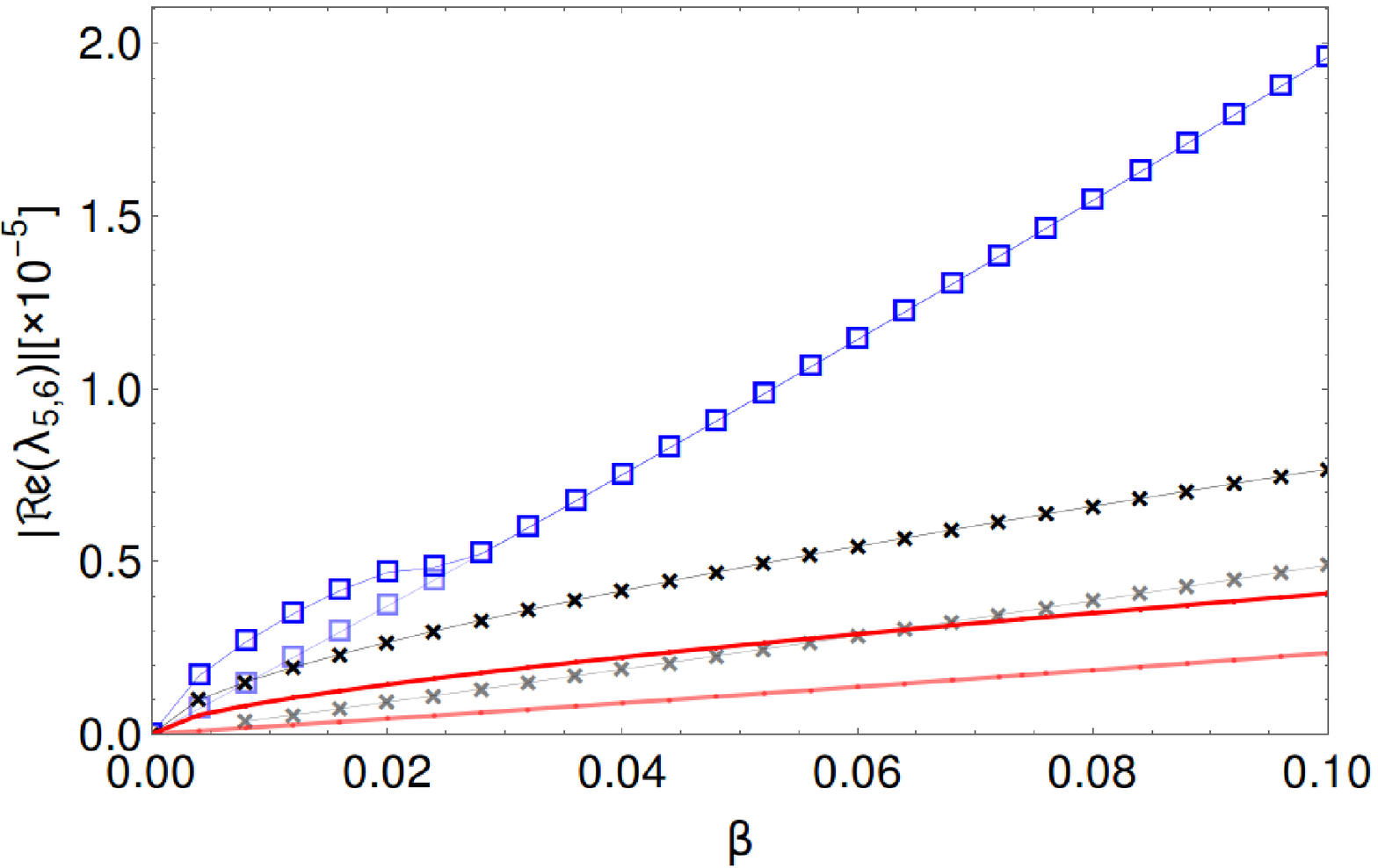}
\includegraphics[width=0.46\linewidth]{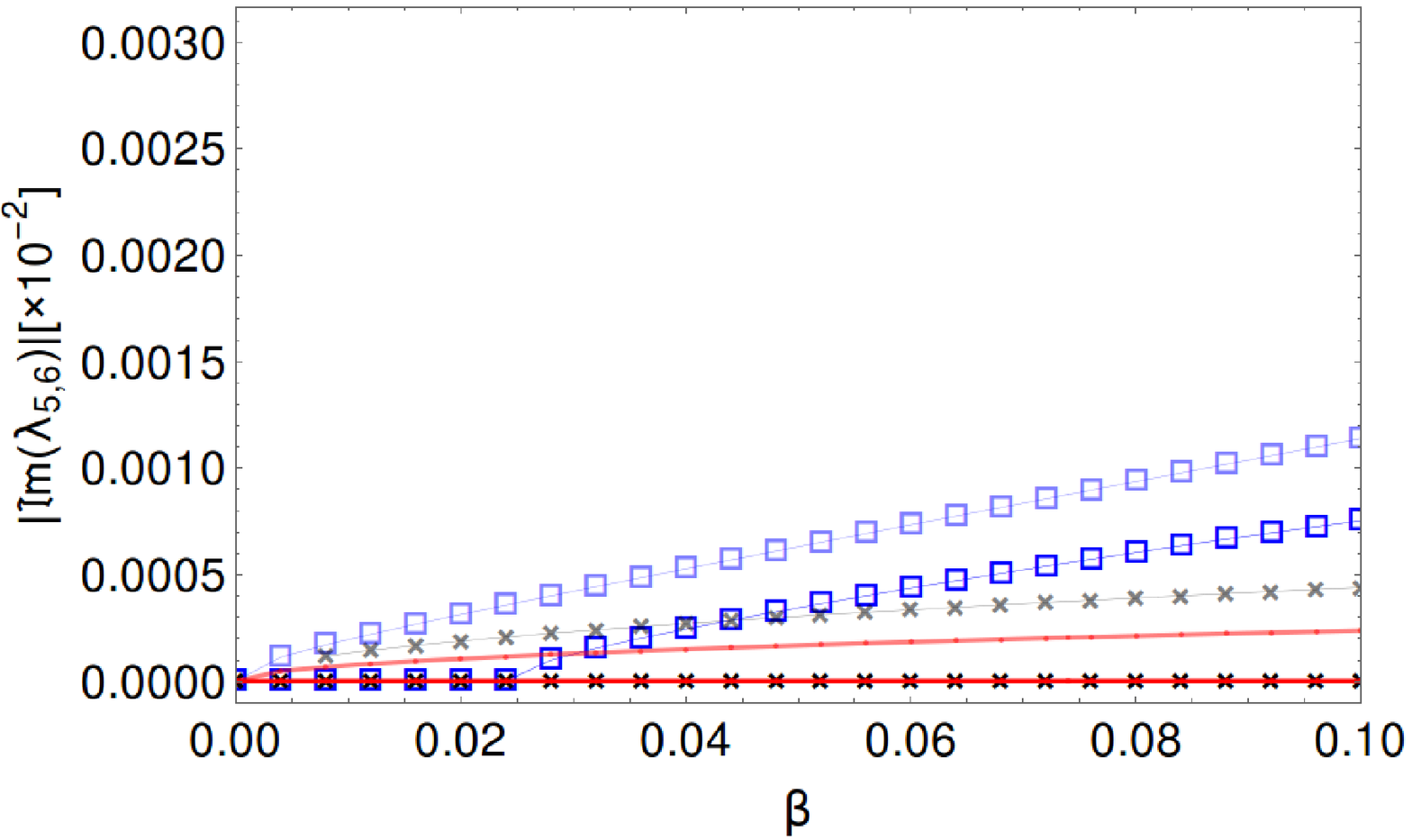}
\caption{Absolute values of real (left column) and imaginary parts
(right column) of $\lambda_{1,2}$ (top row), $\lambda_{3,4}$
(middle row), and $\lambda_{5,6}$ (bottom row) for $e_1=0.1$,
$i_1=5^o$ and for different mass parameters $\mu_1/\mu_J$ equal
$1$ (red dot), $1.5$ (black cross), and $0.6$ (blue square) in
dependency of the dissipative parameter $\beta$. We show the behaviour
close to $L_4$ (thick) and $L_5$ (thin), respectively.}
\label{f:sta}
\end{figure}

In this section we investigate the eigenvalues of the linearized averaged
vector field \equ{disseq} in the neighborhood of the equilibrium. Let us
consider a small displacement close to the equilibrium, say
$(P_0,Q_0,\red{W_0},p_0,q_0,\red{w_0})$:
\beqa{linvar}
P &=& P_0 + \delta P \ , \quad Q = Q_0+\delta Q \ , \quad \red{W = W_0+\delta W} \nonumber \\
p &=& p_0 + \delta p \ , \quad\ \ q = q_0+\delta q \ , \quad\ \ \ \red{w = w_0+\delta w} \ .
\eeqa

The linearization around the equilibrium position provides
the matrix $A=(a_{ij})$ with elements:
\beqano
a_{1\alpha}&=&
-\mu_1\frac{\partial^2 \red{\overline{\mathcal R}}}{\partial p\partial \alpha}
+\frac{\partial Y_P}{\partial \alpha} \ , \quad
a_{4\alpha,\alpha\not=1}=
\mu_1\frac{\partial^2 \red{\overline{\mathcal R}}}{\partial P\partial \alpha} \\
a_{2\alpha}&=&
-\mu_1\frac{\partial^2 \red{\overline{\mathcal R}}}{\partial q\partial \alpha}
+\frac{\partial Y_Q}{\partial \alpha} \ , \quad
a_{5\alpha}=
\mu_1\frac{\partial^2 \red{\overline{\mathcal R}}}{\partial Q\partial \alpha} \\
a_{3\alpha}&=&
-\mu_1\frac{\partial^2 \red{\overline{\mathcal R}}}{\partial{\red{w}}\partial \alpha}
+\frac{\partial{\red{Y_W}}}{\partial \alpha} \ , \quad
a_{6\alpha}=
\mu_1\frac{\partial^2 \red{\overline{\mathcal R}}}{\partial{\red{W}}\partial \alpha}
\eeqano
with $\alpha=P,Q,\red{W},p,q,\red{w}$ and
\beqno
a_{41}=\frac{3 (1-\beta \mu_0)^2}{P^4}
+ \mu_1\frac{\partial^2 \red{\overline{\mathcal R}}}{\partial P} \ .
\eeqno

We immediately notice that all derivatives of $Y_P$ and $Y_Q$ with respect to
$p$, $q$, \red{$w$}, \red{$W$} are zero and that the derivatives of \red{$Y_W$} with respect to $p$, $q$, \red{$w$} are zero.
The solution of the variational equations
\beq{vareq}
\frac{d}{dt}\left(
 \delta P,
 \delta Q,
 \delta \red{W},
 \delta p,
 \delta q,
 \delta \red{w}
\right)^\top
=A\cdot
\left(
 \delta P,
 \delta Q,
 \delta \red{W},
 \delta p,
 \delta q,
 \delta \red{w}
\right)^\top
\eeq

contains terms of the form
\beqno
c_j{\mathbf v}_j e^{\lambda_jt} \ ,
\eeqno

where ($c_j$, ${\mathbf v}_j$, $\lambda_j$), $j=1,...,6$, denotes the
eigensystem of $A$. We show the stability of the linearized tangent flow
on the basis of the eigenvalues
$\lambda_j$ with $j=1,...,6$ in Figure~\ref{f:sta}.  In the top row we show the dependency of
the absolute values of the real and imaginary parts of $\lambda_{1,2}$, related
to the linearized dynamics of the pair $(\delta P, \delta p)$, for varying
$\beta$ and different mass ratios $\mu_1/\mu_J$ from Table~\ref{t:sys}. Our
conclusions are as follows:

\begin{itemize}

\item For $\beta=0$ the real parts are zero for all mass ratios. With increasing
$\beta$ the absolute values of $\lambda_{1,2}$ increase. The slopes are steeper
for larger ratios $\beta/c$, indicating less stable motions.

\item The absolute values of the imaginary parts, that are related to the
fundamental frequencies of motion, decrease with increasing $\beta$, indicating
slightly larger periods of oscillation for larger $\beta$. We also observe,
that larger $a_1$ leads to bigger values of the absolute eigenvalues.

\item We demonstrate that the stability close to $L_4$ is different from
the stability close to $L_5$ as already pointed out in \cite{Mur94} (but based
on the CPRTBP and an oversimplified drag model). \red{Maximum differences in
absolute values between $L_4$ and $L_5$ are largest (left: $10^{-7}$,
right: $10^{-3}$) for the case $\mu_1=0.6$ and much smaller (left: $10^{-9}$,
right: $10^{-4}$) for the cases $\mu_1=1,1.5$.}

\end{itemize}

Next, we investigate the linearized stability of motion of the dynamics related
to the pair of variables $(\delta Q, \delta q)$, see middle row of Figure~\ref{f:sta}:

\begin{itemize}

\item The qualitative behaviour, w.r.t. $\beta$, $\mu_1$, and $a_1$, of the dynamics
of the absolute values of the real and imaginary parts of $\lambda_{3,4}$ is the
same as for the pair  $\lambda_{1,2}$.

\item Instabilities induced in the dynamics of the pair $(\delta Q, \delta q)$
are 100 orders of magnitude stronger than instabilities induced in the dynamics
of the pair $(\delta P, \delta p)$.

\item \red{The maximal differences in absolute values between $L_4$ and $L_5$
are of the order of magnitude of $10^{-9}$ for $\mu_1=1,1.5$ and $10^{-7}$ for
the case $\mu_1=0.6$.}

\end{itemize}

Finally, we find from Figure~\ref{f:sta} (bottom row):

\begin{itemize}

\item The main difference in the linearized motions related to the pair
$(\delta \red{W},\delta \red{w})$ w.r.t. the previous cases is the increase in
oscillation frequency for larger $\beta$ (see bottom, right).

\item \red{The maximum differences in absolute values between $L_4$ and $L_5$
are of the order of $10^{-6}$ for the cases $\mu_1=0.6,1,1.5$.}

\end{itemize}

As a conclusion, due to the presence of non-zero real parts in all pairs of
eigenvalues $\lambda_{i,i+1}$ with $i=1,3,5$, we find an exponential divergence
in the solution of \equ{vareq}; therefore, our system does not provide spectral or
linear stability.  \\

\red{We also notice that the \sl distance \rm of the equilibria in the parameter space for
$\beta\neq0$ from the conservative solution is larger for smaller mass ratios,
and that the difference between $L_4$ and $L_5$ is due to the  asymmetry of nearby
initial conditions - the effect being larger for smaller masses (see further
explanations at the beginning and end of Section~\ref{StaSol}). Moreover, the stability is mainly
affected by the semi-major axis of the perturber: indeed, smaller values of $a_1$ indicate
higher velocities and thus stronger drag terms, that lead to less stable motions, which
correspond to larger absolute values of the real parts in consistency with
Figure~\ref{f:sta}.} \\

\begin{figure}
\centering
\includegraphics[width=0.45\linewidth]{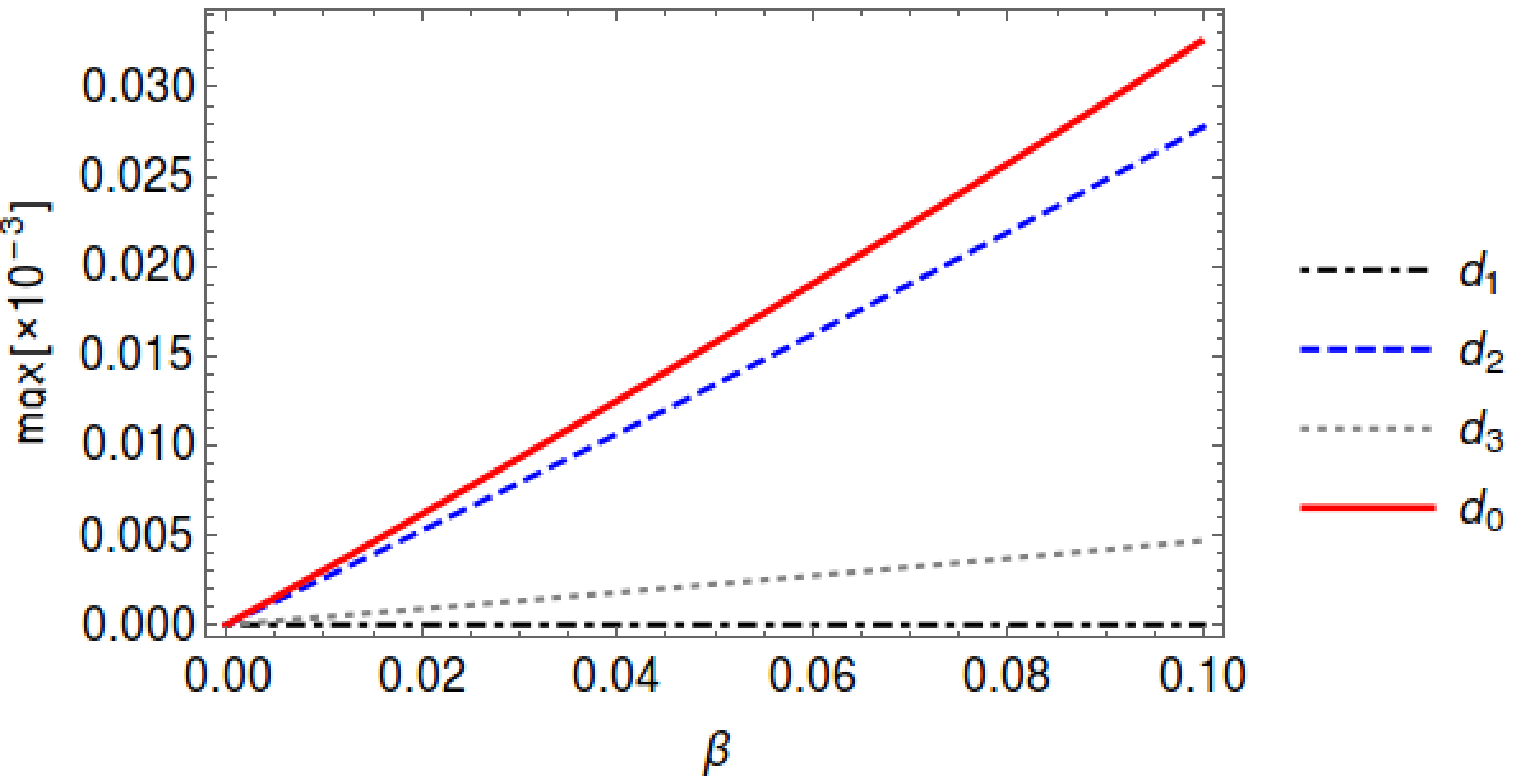}
\includegraphics[width=0.45\linewidth]{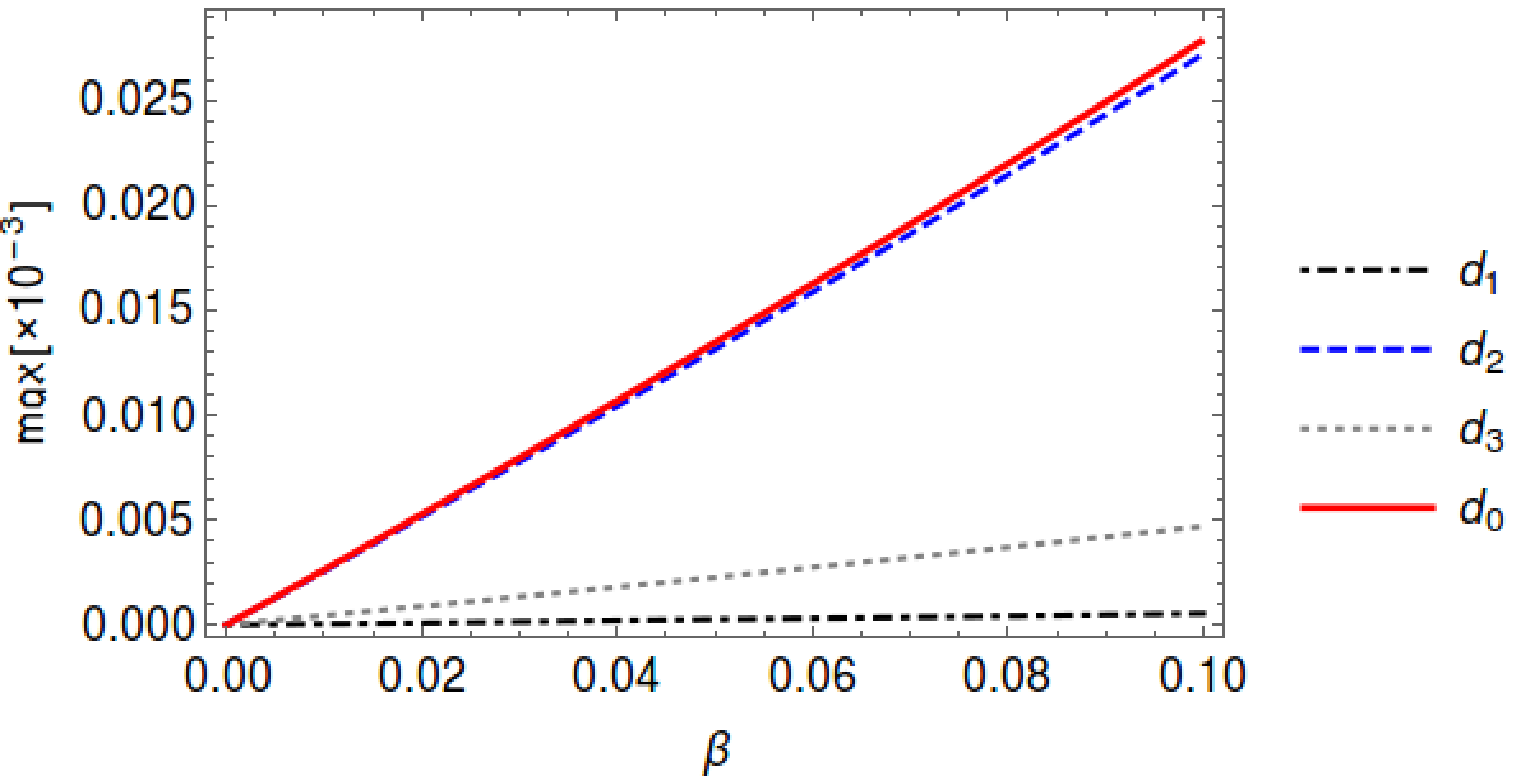}
\caption{Comparison of the first order formulae \equ{eq:d0}-\equ{EVA} on the left
with the numerically obtained figure on the right for $\mu_1=\mu_J$, $e=e_1$,
$i_1=5^{\circ}$.}
\label{f:sta2}
\end{figure}

Finally, we add a remark on the effect of the dissipative parameter $\beta$ on
the symplectic phase space structure. If we denote by $J$ the $6\times 6$
symplectic matrix, it holds for $A$ that:

\beq{ISM}
|\max_{i,j}(A^\top J+JA)_{i,j}|=d_0 \ ,
\eeq
with the maximum computed over all elements of the matrix $A^\top J+JA$ and
with $d_0=0$ only for $\beta=0$. At first order in $\beta$ we find
\beq{eq:d0}
d_0=\frac{\beta}{c}\cdot\frac{7-6e_0^2+\sin^2(i_0/2)}{\sqrt{a_0(1-e_0^2)}}
+ O(\beta^2) \ ,
\eeq

that is proportional to the ratio $\beta/c$ in the same way as the slopes
in the absolute values of the real parts in Figure~\ref{f:sta}. We also notice,
that for $\beta=0$ the sum of the conjugated eigenvalues of $A$
turns out to be zero, because for $\beta=0$ the matrix $A$ becomes an infinitesimally
symplectic matrix. However, for $\beta\not=0$ we find
\beqa{EVA}
d_1&=&|\lambda_1+\lambda_2|=
O(\beta^2) \nonumber \\
d_2&=&|\lambda_3+\lambda_4|=
|\frac{\beta}{c}\cdot\frac{6(e_0^2-1)}{\sqrt{a_0}}| + O(\beta^2) \nonumber \\
d_3&=&|\lambda_5+\lambda_6|=
|\frac{\beta}{c}\cdot\frac{1}{\sqrt{a_0(1-e_0^2)}}| + O(\beta^2) \ .
\eeqa
We remark that, up to first order in $\beta$, the $d_i$'s do not depend on the
mass ratio $\mu_1$. We provide in Figure~\ref{f:sta2} the dependency of $d_i$
versus $\beta$; the left plot is based on our first order formulae
\equ{ISM}--\equ{EVA}, the right plot shows the $d_i$'s obtained as follows. We
calculate the equilibrium values $P_0,Q_0,\red{W_0},p_0,q_0,\red{w_0}$ from \equ{disseq}
for $\mu_1=\mu_J$, $e_1=0.1$, $i_1=5^{\circ}$. Next, we expand the averaged
vector field around the equilibrium to obtain a numerical value for $A$.
Finally, we implement the formula $|\max_{i,j}(A^\top J+JA)_{i,j}|=d_0$,
$|\lambda_1+\lambda_2|$, $|\lambda_3+\lambda_4|$, $|\lambda_5+\lambda_6|$ to
obtain the $d_i$'s in a purely numerical way. As we can see comparing the two
plots of Figure~\ref{f:sta2}, the first order formulae reproduce quite well the
values of the infinitesimally symplectic parameter $d_0$ as well as the values
of $d_1$, $d_2$, $d_3$. \\

\section{Numerical study based on the unaveraged model}\label{NumSur}

\begin{figure}
\centering
\includegraphics[width=0.95\linewidth]{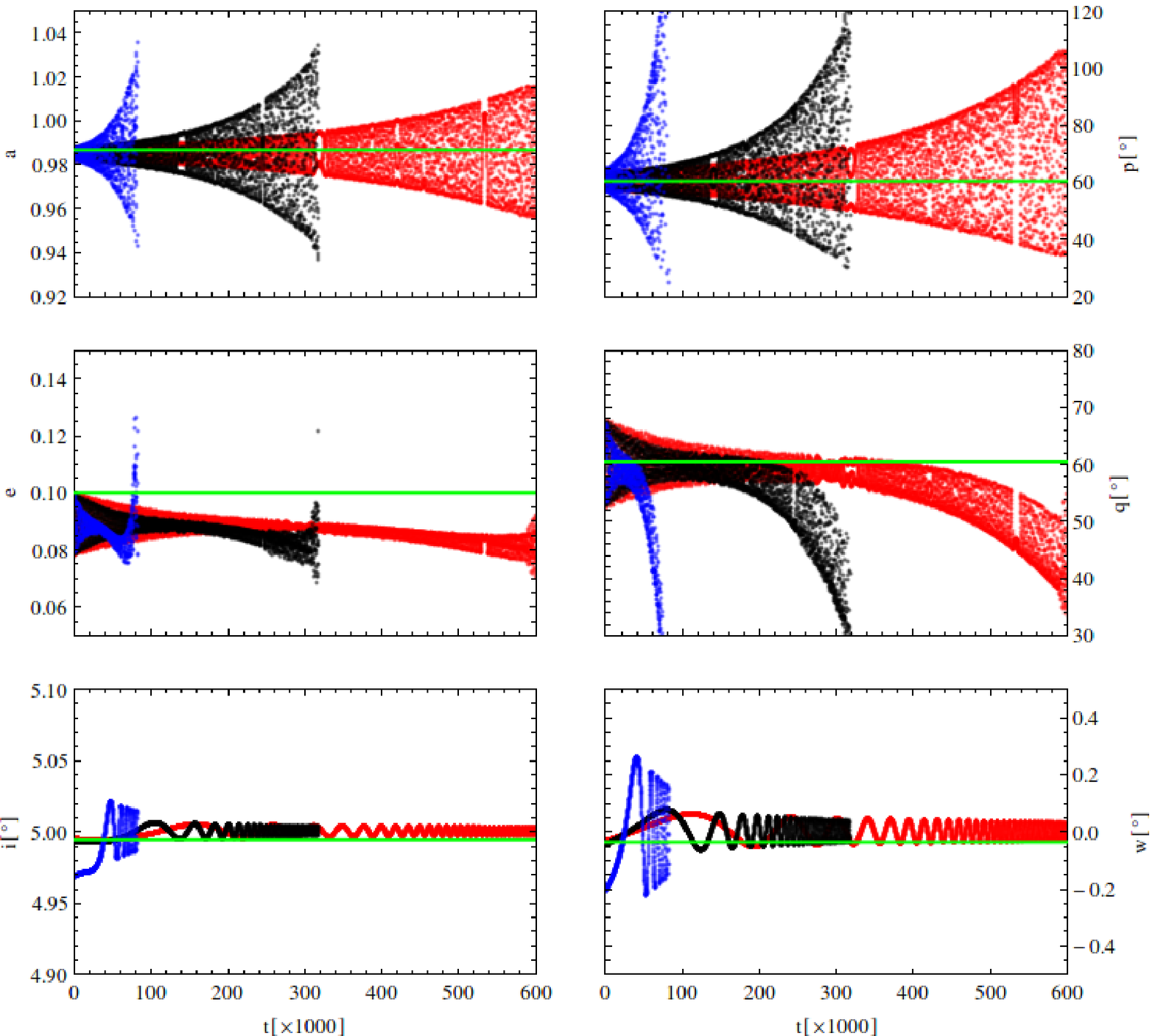}
\caption{Lagrange orbit for $\mu_1/\mu_J$ equal $1$ (red), $1.5$ (black), and
$0.6$ (blue) for $e_1=0.1$, $i_1=5^{\circ}$, and $\beta=0.01$. Initial conditions
coincide with the equilbria of the averaged system (green thick): $a(0)\simeq0.98$,
$e(0)\simeq0.99$, $i(0)\simeq4.99$, $p(0)\simeq60.42^{\circ}$,
$q(0)\simeq60.44^{\circ}$, and $\red{w}(0)\simeq0$.}
\label{n:cross-all}
\end{figure}

In this section we perform a numerical survey to confirm our
results by comparing the analysis of the averaged model with the
unaveraged equations of motion. For this reason we integrate
\equ{eq:vec} using a Runge-Kutta 4-th order integration method
with initial conditions that define the equilibrium of the
averaged dynamics.  We integrate the initial conditions as long
$\psi$ (the angle between $\vec{r}_1$ and $\vec{r})$ stays within
the interval $[10^{\circ},180^{\circ}]$, with a maximum
integration time set to $T=600\,000$ revolution periods of the
secondary.  Since our starting values are obtained from an
averaged model, we expect a slight shift with respect to the
non-averaged model. Moreover, the equilibrium of the averaged
dynamics corresponds to a periodic orbit of the un-averaged
system, that explicitely depends on time $t$ through the
perturbing planet. For the Lagrange orbit associated to $L_4$,
$L_5$ in the SERTBP, we expect for $\beta=0$ that  $a$, $e$, $i$
stay constant, while the resonant angles $p$, $q$ oscillate around
$60^{\circ}$, and the angle \red{$w$}  oscillates around $0^{\circ}$,
respectively. For small $\beta$, say $\beta=0.01$ and different
$\mu_1$, we get a libration of $a$, $p$, $e$, $q$ in
Figure~\ref{n:cross-all}, and an oscillation of $i$, \red{$w$} around
the initial values. For some orbital elements, most notably
semimajor axis and inclination, we observe a small shift between
the averaged motion, that we predicted from averaging theory in
Section~\ref{StaSol}, and the mean value, around which the
elements oscillate, in the unaveraged
dynamics. \\

\begin{figure}
\centering
\includegraphics[width=0.45\linewidth]{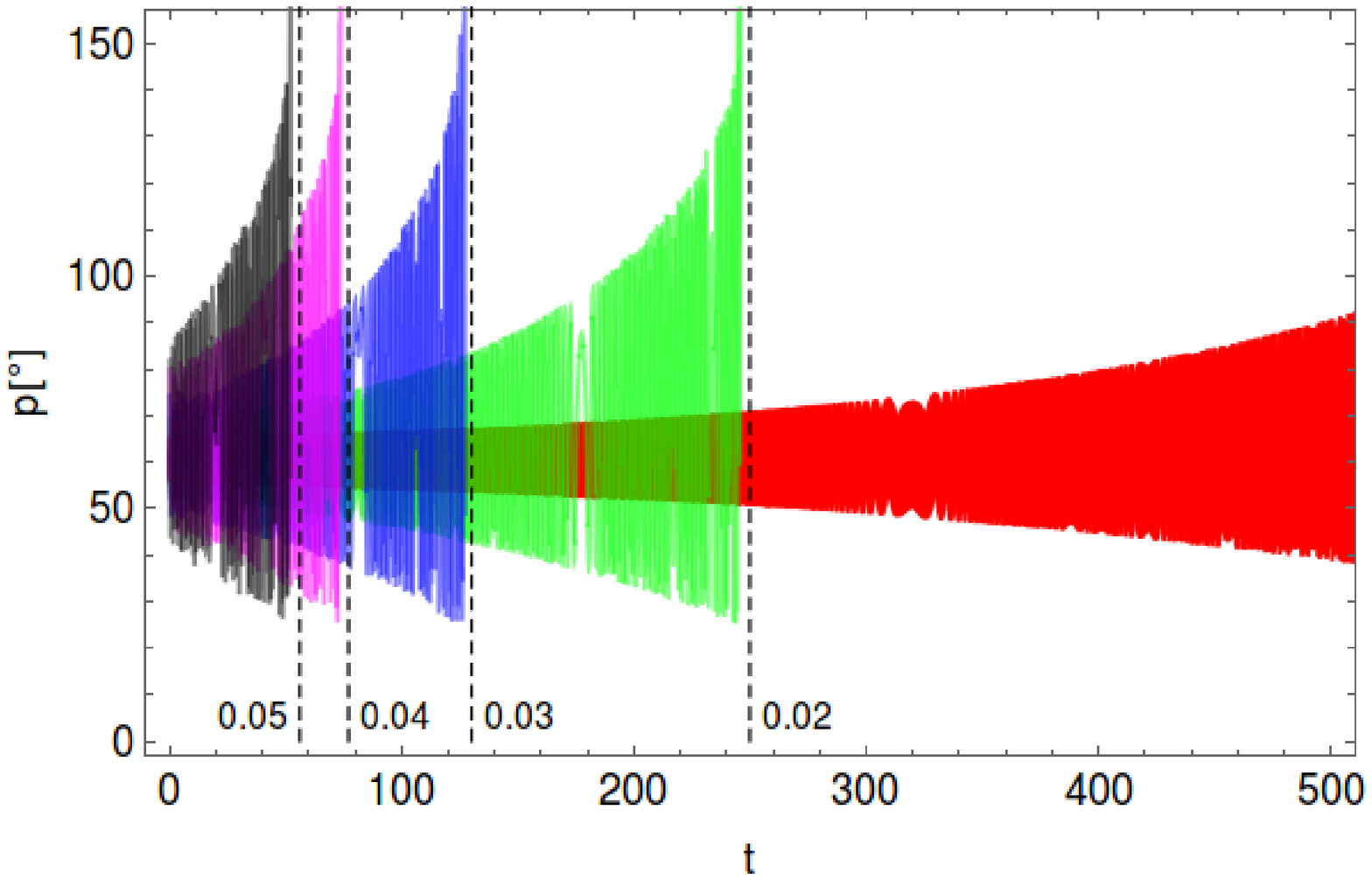}
\includegraphics[width=0.45\linewidth]{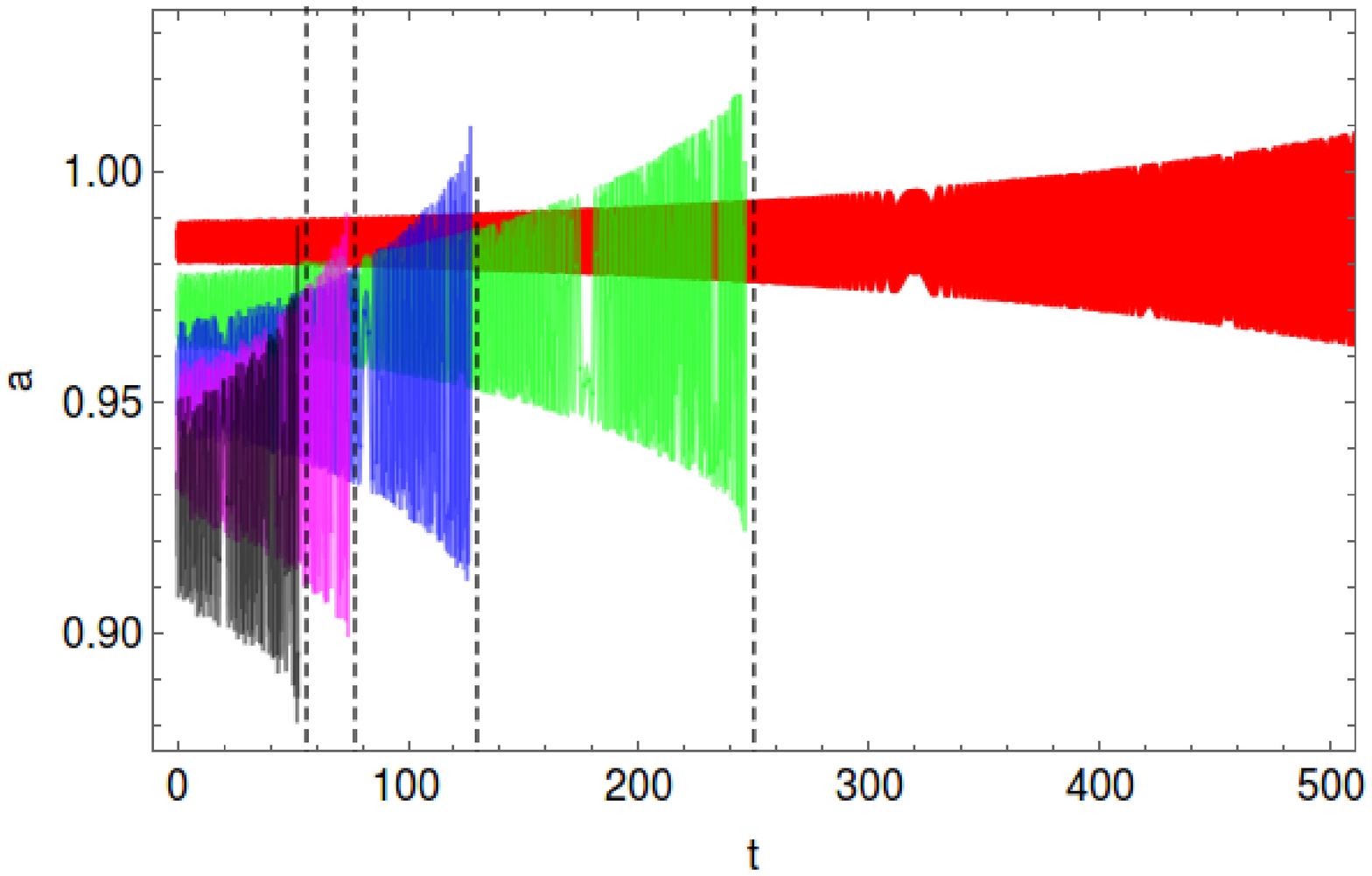}
\caption{The effect of $\beta$ on the time of \red{temporary stability} of
Lagrange orbits for $\mu_1=\mu_J$, $e=e_1$, $i=5^{\circ}$: $0.01$ (red),
$0.02$ (green), $0.03$ (blue), $0.04$ (magenta), and $0.05$ (black).}
\label{n:cross-pP}
\end{figure}

We are left to confirm our prediction of Section~\ref{LinSta},
that dissipative effects act on time-scales proportional to the
ratio $\beta/c$, that is proportional to the quantity $\beta\,
a_1\, n_1$ in arbitrary units: from Figure~\ref{n:cross-all} we
roughly estimate the ratios of the times of \red{temporary stability}
between the Jupiter-like case (red) and case 1 (black) to be about
$2$, and between case 1 and case 2 to be about $4$, that is in
perfect agreement with the values of $\beta\, a_1\, n_1$ that we
may calculate from Table~\ref{t:sys}. We conclude our numerical
survey with a study of the libration width of the elements $p$ and
$a$ in dependency of the parameter $\beta$. We show in
Figure~\ref{n:cross-pP} the evolution in time of the elements $a$
(left) and $p$ (right) for $\beta\in[0.01,0.05]$. As we can see,
for larger values of $\beta$ the element $p$ leaves the
librational resonance earlier, as we already predicted from
averaging theory.

\section{Summary and conclusions}
\label{SumCon}

We investigated the Poynting-Robertson (PR) effect on the co-orbital resonant motion of
dust-sized particles with a planet in the framework of several models, from the circular-planar case to the spatial-elliptic
restricted three-body problem. Our study is based on a simplified resonant
model that we derived on the basis of the equations of motion averaged over
the \red{mean anomaly} of the perturbing planet. We use the resonant model
to find the variation of the equilibrium solution
in the orbital element space of the small particle for different particle size
and mass parameters. We only find \red{temporary stability} of the Lagrange type
orbits in presence of PR drag forces, and show by linear stability analysis
that the instability is due to the steadily increase of the libration width
of the main resonant angle. Our results are validated in several different models
of increasing complexity, and they are confirmed on the basis
of a detailed numerical survey of the unaveraged equations of motion.

The main results of our study are described below.

\begin{itemize}
\item Stable motion for dust sized particles
is not possible due to Poynting-Robertson effect.
\item \red{Temporary stability of particles displaying a tadpole motion} in the
non-averaged system occurs for a wide range of parameters and
initial conditions.
\item The 1:1 resonance with a planet allows a temporary capture of dust size particles
also within the orbit of the perturbing planet, provided it is still in resonance - a fact that has been overseen by
previous studies that found that resonant capture of dust size particles for inner
resonances is not possible due to PR drag.
\item We confirm the presence of a possible asymmetry of the stability indices
of $L_4$ and $L_5$ also in the SERTBP, using a more realistic force model, than it
was used in \cite{Mur94} and based on the CPRTBP.
\end{itemize}

A proper expansion of the perturbing function allows us to
treat the problem by means of averaging theory. The extension of our work to the spatial,
elliptic, restricted three-body (SERTBP) problem shows the importance of the third dimension
in this kind of studies. Inner and outer resonances should therefore be reinvestigated in
the framework of the SERTBP. The effect of dissipative forces on the resonant motion may play
a key role in planetary formation processes. Finally, it would be interesting to investigate the
effect of other dissipative forces on resonant motions.\\

{\bf Acknowledgments} \\

A.C. was partially supported by PRIN-MIUR 2010JJ4KPA$\_$009, GNFM-INdAM and by the European Grant MC-ITN Stardust.
C. L. was financially supported by the Austrian Science Fund (FWF) project J-3206.

\appendix

\section{Basic series expansions used in our study, based on \cite{Stu59}}

Let $J_k$ be the Bessel function of the first kind. The radius $r$ (similar $r_1$)
and its inverse $r^{-1}$ (and $r_1^{-1}$) can be obtained from:
\beqano
\frac{r}{a}&=&1+\frac{1}{2}e^2-2e\sum _{k=1}^{\infty }
\frac{d J_k(k e)}{d e}\frac{\cos (k M)}{k^2} \nonumber\\
\frac{a}{r}&=&1+2\sum _{k=1}^{\infty } J_k(k e)\cos (k M) \ .
\eeqano

The cosine and sine of the true anomaly $f$ are given by:
\beqano
\cos (f)&=&-e+2 \frac{1-e^2}{e}\sum _{k=1}^{\infty }
J_k(k e)\cos (k M) \ , \nonumber \\
\sin (f)&=&2\sqrt{1-e^2}\sum _{k=1}^{\infty }
\frac{d J_k(k e)}{d e}\frac{\sin (k M)}{k} \ .
\eeqano

Let us denote by $\xi$, $\eta$, $\zeta$ the position of a celestial body
in the orbital frame (where $\zeta=0$). In this setting we have
\beqano
\frac{\xi}{a}&=&-\frac{3}{2}e+2\sum _{k=1}^{\infty } \frac{d J_k(k e)}{d e}
\frac{\cos (k M)}{k^2} \ , \nonumber \\
\frac{\eta}{a}&=&2\frac{\sqrt{1-e^2}}{e}\sum _{k=1}^{\infty }
J_k(k e)\frac{\sin (k M)}{k} \ . \nonumber \\
\eeqano

Time derivatives $\dot r$, $\dot \xi$, $\dot \eta$, $\dot \zeta$ can be directly
obtained from $d/dt$ (assuming $M=n t$):
\beqano
\frac{\dot{r}}{a}&=&2 n e\sum _{\nu =1}^{\infty }
\frac{d J_{\nu }(\nu  e)}{d e}\frac{\sin (M)}{\nu }  \ , \nonumber \\
\frac{\dot \xi}{a}&=&-2n\sum _{k=1}^{\infty } \frac{d J_k(k e)}{d e}
\frac{\sin (k M)}{k} \ , \nonumber \\
\frac{\dot \eta}{a}&=&2n\frac{\sqrt{1-e^2}}{e}\sum _{k=1}^{\infty }
J_k(k e)\cos (k M) \ , \nonumber \ .
\eeqano

The transformation to the inertial reference frame is given
by the rotation matrix $R_M=R_3(\Omega)R_1(i)R_3(\omega)$,
where $R_i$ denotes the rotation around the $i$-th axis ($x$, $y$, $z$). Using the notation $c_\#=\cos(\#)$, $s_\#=\sin(\#)$ we find:
\beqno
R_M=\left(
\begin{array}{ccc}
 c_\omega c_\Omega - c_i s_\omega d_\Omega &
s_\omega - c_\Omega - c_i c_\omega s_\Omega &
s_i s_\Omega \\
 c_i s_\omega c_\Omega + c_\omega s_\Omega &
c_i c_\omega c_\Omega - s_\omega s_\Omega &
s_i - c_\Omega \\
 s_i s_\omega &
s_i c_\omega &
c_i \\
\end{array}
\right) .
\eeqno


\section{Equilateral perturbing function}

We start from the expression of ${\mathcal R}$ given in \equ{PFrr1psi}. Using the small parameter $\rho=\frac{r}{r_1}-1$,
the distance $\Delta^{-1}$ becomes in terms of $\rho$:
\beqno
\frac{1}{\Delta}=
\frac{1}{r_1}\frac{1}{\sqrt{A + A \rho +\rho ^2}} \ .
\eeqno

with $A=2\left(1-\cos\psi\right)$. Setting $\epsilon=\rho+\frac{\rho^2}{A}$ we find
\beqno
\red{\frac{1}{\Delta}}=\frac{1}{r_1}
\frac{1}{\sqrt{2}}
\frac{1}{\sqrt{1-\cos\psi}}
\frac{1}{\sqrt{1+\epsilon}} \ ,
\eeqno

provided $|\cos\psi|<1$, $|\epsilon|<1$; the expansion of $\Delta^{-1}$ becomes
\beqno
\Delta ^{-1}\simeq
\frac{1}{\sqrt{2}}\frac{1}{r_1}
\sum _{j=0}^\infty
\red{\left(-1\right)^j
\left(
\begin{array}{c}
 -1/2 \\
 j \\
\end{array}
\right) \cos (\psi)^j} \sum _{n=0}^\infty
\left(
\begin{array}{c}
 -1/2 \\
 n \\
\end{array}
\right)
\epsilon^n \ ,
\eeqno

and the expansion of the perturbing function, valid close to $r/r_1\simeq 1$, is given by
\beqano
{\mathcal R}=
\frac{1}{\sqrt{2}}\frac{1}{r_1}
\sum _{j=0}^\infty
\red{\left(-1\right)^j
\left(
\begin{array}{c}
 -1/2 \\
 j \\
\end{array}
\right) \cos (\psi)^j} \sum _{n=0}^\infty
\left(
\begin{array}{c}
 -1/2 \\
 n \\
\end{array}
\right)
\epsilon^n
-\frac{r \cos\psi}{r_1^2}
\red{- \frac{1}{r}} \ .
\eeqano
If we compute the expansion up to the order $2$ in $\rho$ and order $2$ in $\cos\psi$, we get the expression
\equ{PF11}. Setting \red{$\alpha=a/a_1-1$}, $\cos i=1-s^2$, $\sin i=2s$,
$\lambda=M+\tilde\omega$, $\tilde\omega=\omega+\Omega$ (and analogously
for $s_1$, $\lambda_1$, $\tilde\omega_1$), and using standard series
expansions for $r$, $r_1$, and $\cos(\psi)$ we find:

\beqano
&&\red{\frac{1}{r}}+a_1 {\mathcal R}=-\frac{10 \alpha ^2+152 \alpha +112}{256 \sqrt{2}}
-\frac{\left(15 \alpha^2+96 \alpha +81\right) e^2}{256 \sqrt{2}}- \\
&&\frac{3 \left(5 \alpha ^2+32\alpha +27\right) e_1^2}{256 \sqrt{2}}
-\frac{\left(-84 \alpha ^2-48\alpha +96\right) s^2}{256 \sqrt{2}}
+\frac{3 \left(7 \alpha ^2+4 \alpha-8\right) s_1^2}{64 \sqrt{2}} \\
&&+\left(\frac{5 \alpha ^2}{64 \sqrt{2}}+\frac{43 \alpha }{64
   \sqrt{2}}+\frac{19}{32 \sqrt{2}}\right) e \cos \left(\lambda
   -\tilde{\omega }\right) \\
&&+\cos \left(\lambda -\lambda _1\right)\bigg(
-\frac{3 \alpha ^2}{16\sqrt{2}}
+\left(-1-\frac{1}{4 \sqrt{2}}\right) \alpha \\
&&+\left(-\frac{3\alpha ^2}{32 \sqrt{2}}
+\left(\frac{1}{2}-\frac{1}{4 \sqrt{2}}\right)\alpha
-\frac{23}{32 \sqrt{2}}+\frac{1}{2}\right) e^2 \\
&&+\left(-\frac{3\alpha ^2}{32 \sqrt{2}}
+\left(\frac{1}{2}-\frac{1}{4 \sqrt{2}}\right)\alpha
-\frac{23}{32 \sqrt{2}}+\frac{1}{2}\right) e_1^2 \\
&&+\left(\frac{3\alpha ^2}{16 \sqrt{2}}
+\left(1+\frac{1}{4 \sqrt{2}}\right) \alpha
-\frac{1}{2 \sqrt{2}}+1\right) s^2 \\
&&+\left(\frac{3 \alpha ^2}{16\sqrt{2}}
+\left(1+\frac{1}{4 \sqrt{2}}\right) \alpha
-\frac{1}{2\sqrt{2}}+1\right) s_1^2
+\frac{1}{2 \sqrt{2}}-1
\bigg) \\
&&+\left(-\frac{21 \alpha ^2}{32 \sqrt{2}}-\frac{3 \alpha }{8
   \sqrt{2}}+\frac{3}{4 \sqrt{2}}\right) s s_1 \cos \left(\Omega -\Omega
   _1\right) \\
&&+\left(\frac{3 \alpha ^2}{16 \sqrt{2}}+\frac{3 \alpha }{8
   \sqrt{2}}+\frac{15}{32 \sqrt{2}}\right) e e_1 \cos \left(\tilde{\omega
   }-\tilde{\omega }_1\right) \\
&&+\left(-\frac{3 \alpha ^2}{32 \sqrt{2}}-\frac{3 \alpha }{16
   \sqrt{2}}-\frac{3}{8 \sqrt{2}}\right) e_1 \cos \left(\lambda
   -\tilde{\omega }_1\right) \\
&&+\left(\frac{3 \alpha ^2}{8 \sqrt{2}}+\left(\frac{3}{2}+\frac{9}{16
   \sqrt{2}}\right) \alpha -\frac{3}{8 \sqrt{2}}+\frac{3}{2}\right) e \cos
   \left(\tilde{\omega }-\lambda _1\right) \\
&&+\left(-\frac{15 \alpha ^2}{128 \sqrt{2}}-\frac{81 \alpha }{64
   \sqrt{2}}-\frac{33}{32 \sqrt{2}}\right) e_1 \cos \left(\lambda
   _1-\tilde{\omega }_1\right) \\
&&+\left(-\frac{15 \alpha ^2}{128 \sqrt{2}}+\left(-\frac{1}{8}-\frac{7}{32
   \sqrt{2}}\right) \alpha -\frac{11}{64 \sqrt{2}}-\frac{1}{8}\right) e^2
   \cos \left(-2 \tilde{\omega }+\lambda +\lambda _1\right) \\
\eeqano
\beqano
&&+\left(-\frac{3 \alpha ^2}{16 \sqrt{2}}+\left(-1-\frac{1}{4 \sqrt{2}}\right)
   \alpha +\frac{1}{2 \sqrt{2}}-1\right) s^2 \cos \left(\lambda +\lambda
   _1-2 \Omega \right) \\
&&+\left(\frac{63 \alpha ^2}{128 \sqrt{2}}+\frac{51 \alpha }{128
   \sqrt{2}}-\frac{21}{64 \sqrt{2}}\right) e \cos \left(\tilde{\omega
   }+\lambda -2 \lambda _1\right) \\
&&+\left(-\frac{3 \alpha ^2}{16 \sqrt{2}}+\left(-1-\frac{1}{4 \sqrt{2}}\right)
   \alpha +\frac{1}{2 \sqrt{2}}-1\right) s_1^2 \cos \left(\lambda +\lambda
   _1-2 \Omega _1\right) \\
&&+\left(-\frac{3 \alpha ^2}{8 \sqrt{2}}+\left(-2-\frac{1}{2 \sqrt{2}}\right)
   \alpha +\frac{1}{\sqrt{2}}-2\right) s s_1 \cos \left(\lambda -\lambda
   _1-\Omega +\Omega _1\right) \\
&&+\left(\frac{21 \alpha ^2}{32 \sqrt{2}}+\frac{3 \alpha }{8
   \sqrt{2}}-\frac{3}{4 \sqrt{2}}\right) s s_1 \cos \left(-2 \lambda
   _1+\Omega +\Omega _1\right) \\
&&+\left(\frac{3 \alpha ^2}{8 \sqrt{2}}+\left(2+\frac{1}{2 \sqrt{2}}\right)
   \alpha -\frac{1}{\sqrt{2}}+2\right) s s_1 \cos \left(\lambda +\lambda
   _1-\Omega -\Omega _1\right) \\
&&+\left(-\frac{15 \alpha ^2}{128 \sqrt{2}}+\left(-\frac{1}{8}-\frac{7}{32
   \sqrt{2}}\right) \alpha -\frac{11}{64 \sqrt{2}}-\frac{1}{8}\right) e_1^2
   \cos \left(-2 \tilde{\omega }_1+\lambda +\lambda _1\right) \\
&&+\left(\frac{15 \alpha ^2}{128 \sqrt{2}}+\frac{3 \alpha }{4
   \sqrt{2}}+\frac{81}{128 \sqrt{2}}\right) e e_1 \cos \left(-\tilde{\omega
   }+\tilde{\omega }_1+\lambda -\lambda _1\right) \\
&&+\left(\frac{15 \alpha ^2}{16 \sqrt{2}}+\left(3+\frac{3}{2 \sqrt{2}}\right)
   \alpha -\frac{9}{32 \sqrt{2}}+3\right) e e_1 \cos \left(\tilde{\omega
   }+\tilde{\omega }_1-2 \lambda _1\right) \\
&&+\left(-\frac{15 \alpha ^2}{32 \sqrt{2}}+\left(-2-\frac{11}{16
   \sqrt{2}}\right) \alpha +\frac{5}{8 \sqrt{2}}-2\right) e_1 \cos
   \left(\tilde{\omega }_1+\lambda -2 \lambda _1\right) \\
&&+\left(\frac{15 \alpha ^2}{128 \sqrt{2}}+\frac{3 \alpha }{4
   \sqrt{2}}+\frac{81}{128 \sqrt{2}}\right) e e_1 \cos \left(-\tilde{\omega
   }-\tilde{\omega }_1+\lambda +\lambda _1\right) \\
&&+\left(-\frac{63 \alpha ^2}{256 \sqrt{2}}+\frac{3 \alpha }{32
   \sqrt{2}}+\frac{177}{256 \sqrt{2}}\right) e e_1 \cos \left(\tilde{\omega
   }-\tilde{\omega }_1+\lambda -\lambda _1\right) \\
&&+\left(\frac{5 \alpha ^2}{256 \sqrt{2}}+\frac{19 \alpha }{64
   \sqrt{2}}+\frac{71}{256 \sqrt{2}}\right) e^2 \cos \left(2 \lambda -2
   \tilde{\omega }\right) \\
&&+\left(-\frac{21 \alpha ^2}{64 \sqrt{2}}-\frac{3 \alpha }{16
   \sqrt{2}}+\frac{3}{8 \sqrt{2}}\right) s^2 \cos (2 \lambda -2 \Omega ) \\
&&+\cos \left(2 \lambda -2 \lambda _1\right) \bigg(
-\frac{21 \alpha ^2}{128\sqrt{2}}
-\frac{3 \alpha }{32 \sqrt{2}}
+\left(\frac{105 \alpha ^2}{256\sqrt{2}}
-\frac{225}{256 \sqrt{2}}\right) e^2 \\
&&+\left(\frac{105 \alpha^2}{256 \sqrt{2}}
-\frac{225}{256 \sqrt{2}}\right) e_1^2
+\left(\frac{21\alpha ^2}{64 \sqrt{2}}
+\frac{3 \alpha }{16 \sqrt{2}}
-\frac{3}{8\sqrt{2}}\right) s^2 \\
\eeqano
\beqano
&&+\left(\frac{21 \alpha ^2}{64 \sqrt{2}}
+\frac{3\alpha }{16 \sqrt{2}}
-\frac{3}{8 \sqrt{2}}\right) s_1^2
+\frac{3}{16\sqrt{2}}\bigg) \\
&&+\left(-\frac{21 \alpha ^2}{64 \sqrt{2}}-\frac{3 \alpha }{16
   \sqrt{2}}+\frac{3}{8 \sqrt{2}}\right) s_1^2 \cos \left(2 \lambda -2
   \Omega _1\right) \\
&&+\left(\frac{21 \alpha ^2}{32 \sqrt{2}}+\frac{3 \alpha }{8
   \sqrt{2}}-\frac{3}{4 \sqrt{2}}\right) s s_1 \cos \left(2 \lambda -\Omega
   -\Omega _1\right) \\
&&+\left(\frac{441 \alpha ^2}{256 \sqrt{2}}+\frac{27 \alpha }{16 \sqrt{2}}-\frac{159}{256 \sqrt{2}}\right) e e_1 \cos \left(\tilde{\omega }+\tilde{\omega }_1+\lambda -3 \lambda _1\right)+\frac{15 e_1^2 \cos \left(2 \lambda -2 \tilde{\omega }_1\right)}{512 \sqrt{2}} \\
&&+\left(\left(\frac{1}{16 \sqrt{2}}-\frac{1}{2}\right) \alpha +\frac{5}{8 \sqrt{2}}-\frac{1}{2}\right) e \cos \left(-\tilde{\omega }+2 \lambda -\lambda _1\right)+-\frac{9 e e_1 \cos \left(-\tilde{\omega }-\tilde{\omega }_1+2 \lambda \right)}{32 \sqrt{2}} \\
&&+\left(-\frac{21 \alpha ^2}{64 \sqrt{2}}-\frac{3 \alpha }{16 \sqrt{2}}
+\frac{3}{8 \sqrt{2}}\right) s^2 \cos \left(2 \Omega -2 \lambda _1\right) \\
&&+\left(-\frac{105 \alpha ^2}{256 \sqrt{2}}-\frac{15 \alpha }{32 \sqrt{2}}+\frac{15}{512 \sqrt{2}}\right) e^2 \cos \left(2 \tilde{\omega }-2 \lambda _1\right) \\
&&+\left(-\frac{21 \alpha ^2}{64 \sqrt{2}}-\frac{3 \alpha }{16
   \sqrt{2}}+\frac{3}{8 \sqrt{2}}\right) s_1^2 \cos \left(2 \lambda _1-2
   \Omega _1\right) \\
&&+\left(-\frac{45 \alpha ^2}{256 \sqrt{2}}-\frac{105 \alpha }{64
   \sqrt{2}}-\frac{345}{256 \sqrt{2}}\right) e_1^2 \cos \left(2 \lambda _1-2
   \tilde{\omega }_1\right) \\
&&+\left(\frac{21 \alpha ^2}{256 \sqrt{2}}-\frac{9 \alpha }{128
   \sqrt{2}}-\frac{21}{64 \sqrt{2}}\right) e_1 \cos \left(-\tilde{\omega
   }_1+2 \lambda -\lambda _1\right) \\
&&+\left(-\frac{21 \alpha ^2}{32 \sqrt{2}}-\frac{3 \alpha }{8
   \sqrt{2}}+\frac{3}{4 \sqrt{2}}\right) s s_1 \cos \left(2 \lambda -2
   \lambda _1-\Omega +\Omega _1\right) \\
&&+\left(-\frac{117 \alpha ^2}{128 \sqrt{2}}+\left(-\frac{27}{8}-\frac{45}{32
   \sqrt{2}}\right) \alpha +\frac{45}{64 \sqrt{2}}-\frac{27}{8}\right) e_1^2
   \cos \left(2 \tilde{\omega }_1+\lambda -3 \lambda _1\right) \\
&&+\left(\left(\frac{1}{8 \sqrt{2}}-1\right) \alpha +\frac{31}{32
   \sqrt{2}}-1\right) e e_1 \cos \left(-\tilde{\omega }+\tilde{\omega }_1+2
   \lambda -2 \lambda _1\right) \\
&&+\left(-\frac{147 \alpha ^2}{256 \sqrt{2}}-\frac{57 \alpha }{128
   \sqrt{2}}+\frac{27}{64 \sqrt{2}}\right) e_1 \cos \left(\tilde{\omega
   }_1+2 \lambda -3 \lambda _1\right) \\
&&+\left(\frac{3 \alpha ^2}{128 \sqrt{2}}+\left(\frac{3}{32
   \sqrt{2}}-\frac{3}{8}\right) \alpha +\frac{45}{64
   \sqrt{2}}-\frac{3}{8}\right) e^2 \cos \left(-2 \tilde{\omega }+3 \lambda
   -\lambda _1\right) \\
&&+\left(-\frac{21 \alpha ^2}{128 \sqrt{2}}+\frac{3 \alpha }{128
   \sqrt{2}}+\frac{27}{64 \sqrt{2}}\right) e \cos \left(-\tilde{\omega }+3
   \lambda -2 \lambda _1\right) \\
&&+\left(\frac{21 \alpha ^2}{256 \sqrt{2}}-\frac{3 \alpha }{16
   \sqrt{2}}-\frac{159}{256 \sqrt{2}}\right) e e_1 \cos \left(-\tilde{\omega
   }-\tilde{\omega }_1+3 \lambda -\lambda _1\right) \\
\eeqano
\beqano
&&+\left(-\frac{147 \alpha ^2}{256 \sqrt{2}}-\frac{3 \alpha }{32
   \sqrt{2}}+\frac{273}{256 \sqrt{2}}\right) e e_1 \cos \left(-\tilde{\omega
   }+\tilde{\omega }_1+3 \lambda -3 \lambda _1\right) \\
&&+\left(-\frac{21 \alpha ^2}{128 \sqrt{2}}+\frac{9 \alpha }{64
   \sqrt{2}}+\frac{351}{512 \sqrt{2}}\right) e^2 \cos \left(-2 \tilde{\omega
   }+4 \lambda -2 \lambda _1\right) \\
&&+\left(-\frac{357 \alpha ^2}{256 \sqrt{2}}-\frac{81 \alpha }{64
   \sqrt{2}}+\frac{351}{512 \sqrt{2}}\right) e_1^2 \cos \left(2
   \tilde{\omega }_1+2 \lambda -4 \lambda _1\right) \ ,
\eeqano

\red{where the term $1/r$ is given by}
\beqno
\red{
\frac{1}{r}=
\frac{
1
-e\cos\left(\lambda-\tilde{\omega}\right)
+e^2\cos\left(2\lambda-2\tilde{\omega}\right)}
{\left(1+\alpha\right)}
}
\eeqno

We notice that for our study we used the orders $8$ in $\rho$ and $24$
in $\psi$ to obtain $\mathcal R$. \red{These expansions are necessary
to ensure that the difference between \equ{PFrr1psi} and its expansion
is less than the machine precision close to the equilibrium points
$L_4$ and $L_5$.} \\

\bibliographystyle{elsarticle-harv}
\bibliography{biblio}

\begin{thebibliography}{32}
\expandafter\ifx\csname natexlab\endcsname\relax\def\natexlab#1{#1}\fi
\expandafter\ifx\csname url\endcsname\relax
  \def\url#1{\texttt{#1}}\fi
\expandafter\ifx\csname urlprefix\endcsname\relax\def\urlprefix{URL }\fi

\bibitem[{{Beaug\'e} and {Ferraz-Mello}(1994)}]{Bea94}
{Beaug\'e}, C., {Ferraz-Mello}, S., Aug. 1994. {Capture in exterior mean-motion
  resonances due to Poynting-Robertson drag}. Icarus 110, 239--260.

\bibitem[{{Brown} and {Shook}(1964)}]{BroSho64}
{Brown}, E., {Shook}, C., 1964. Planetary Theory. Dover books on astronomy and
  astrophysics, 1133.

\bibitem[{{Burns} et~al.(1979){Burns}, {Lamy}, and {Soter}}]{Buretal79}
{Burns}, J., {Lamy}, P., {Soter}, S., 1979. Radiation forces on small particles
  in the solar system. Icarus 40, 1--48.

\bibitem[{{Burns} et~al.(2014){Burns}, {Lamy}, and {Soter}}]{Bur14}
{Burns}, J.~A., {Lamy}, P.~L., {Soter}, S., Apr. 2014. {Radiation forces on
  small particles in the Solar System: A re-consideration}. Icarus 232,
  263--265.

\bibitem[{Celletti(2010)}]{alebook}
Celletti, A., 2010. Stability and chaos in celestial mechanics.
  Springer-Verlag, Berlin; published in association with Praxis Publishing
  Ltd., Chichester.
\newline\urlprefix\url{http://dx.doi.org/10.1007/978-3-540-85146-2}

\bibitem[{{Das} et~al.(2008){Das}, {Narang}, {Mahajan}, and {Yuasa}}]{Das08}
{Das}, M.~K., {Narang}, P., {Mahajan}, S., {Yuasa}, M., Apr. 2008. {Effect of
  radiation on the stability of equilibrium points in the binary stellar
  systems: RW-Monocerotis, Kr{\"u}ger 60}. Astrophysics \& Space Science 314,
  261--274.

\bibitem[{{Dermott} et~al.(1994){Dermott}, {Jayaraman}, {Xu}, {Gustafson}, and
  {Liou}}]{Der94}
{Dermott}, S.~F., {Jayaraman}, S., {Xu}, Y.~L., {Gustafson}, B.~{\AA}.~S.,
  {Liou}, J.~C., Jun. 1994. {A circumsolar ring of asteroidal dust in resonant
  lock with the Earth}. Nature 369, 719--723.

\bibitem[{Dvorak and Lhotka(2013)}]{chrbook}
Dvorak, R., Lhotka, C., 2013. Celestial Dynamics. WILEY.

\bibitem[{{Espy} et~al.(2008){Espy}, {Dermott}, and {Kehoe}}]{Esp08}
{Espy}, A.~J., {Dermott}, S.~F., {Kehoe}, T.~J.~J., Jun. 2008. {Dynamical
  Effects of Mars on Asteroidal Dust Particles}. Earth Moon and Planets 102,
  199--203.

\bibitem[{{Gr\"un} et~al.(1985){Gr\"un}, {Zook}, {Fechtig}, and
  {Giese}}]{Gru85}
{Gr\"un}, E., {Zook}, H.~A., {Fechtig}, H., {Giese}, R.~H., May 1985.
  {Collisional balance of the meteoritic complex}. Icarus 62, 244--272.

\bibitem[{{Gustafson}(1994)}]{Gus94}
{Gustafson}, B.~A.~S., 1994. {Physics of Zodiacal Dust}. Annual Review of Earth
  and Planetary Sciences 22, 553--595.

\bibitem[{Jancart and Lemaitre(2001)}]{JL}
Jancart, S., Lemaitre, A., 2001. Dissipative forces and external resonances.
  Celestial Mech. Dynam. Astronom. 81~(1-2), 75--80, dynamics of natural and
  artificial celestial bodies (Pozna{\'n}, 2000).
\newline\urlprefix\url{http://dx.doi.org/10.1023/A:1013311204539}

\bibitem[{{Kla{\v c}ka}(2013)}]{Kla13}
{Kla{\v c}ka}, J., Dec. 2013. {Comparison of the solar/stellar wind and the
  Poynting-Robertson effect in secular orbital evolution of dust particles}.
  MNRAS 436, 2785--2792.

\bibitem[{{Kla{\v c}ka}(2014)}]{Kla14b}
{Kla{\v c}ka}, J., 2014. {Solar wind dominance over the Poynting-Robertson
  effect in secular orbital evolution of dust particles}. MNRAS 443, 213--229.

\bibitem[{{Kla{\v c}ka} and {Kocifaj}(2008)}]{Kla08a}
{Kla{\v c}ka}, J., {Kocifaj}, M., Nov. 2008. {Times of inspiralling for
  interplanetary dust grains}. MNRAS 390, 1491--1495.

\bibitem[{{Kla{\v c}ka} et~al.(2008){Kla{\v c}ka}, {K{\'o}mar}, {P{\'a}stor},
  and {Petr{\v z}ala}}]{Kla08b}
{Kla{\v c}ka}, J., {K{\'o}mar}, L., {P{\'a}stor}, P., {Petr{\v z}ala}, J., Oct.
  2008. {The non-radial component of the solar wind and motion of dust near
  mean motion resonances with planets}. A\&A 489, 787--793.

\bibitem[{{Kla{\v c}ka} et~al.(2014){Kla{\v c}ka}, {Petr{\v z}ala},
  {P{\'a}stor}, and {K{\'o}mar}}]{Kla14}
{Kla{\v c}ka}, J., {Petr{\v z}ala}, J., {P{\'a}stor}, P., {K{\'o}mar}, L., Apr.
  2014. {The Poynting-Robertson effect: A critical perspective}. Icarus 232,
  249--262.

\bibitem[{{Kocifaj} and {Kla{\v c}ka}(2008)}]{Koc08}
{Kocifaj}, M., {Kla{\v c}ka}, J., May 2008. {Nonspherical dust grains in
  mean-motion orbital resonances}. A\&A 483, 311--315.

\bibitem[{{Kocifaj} and {Kundracik}(2012)}]{Koc12}
{Kocifaj}, M., {Kundracik}, F., May 2012. {On some microphysical properties of
  dust grains captured into resonances with Neptune}. MNRAS 422, 1665--1673.

\bibitem[{{Kortenkamp}(2013)}]{Kor13}
{Kortenkamp}, S.~J., Nov. 2013. {Trapping and dynamical evolution of
  interplanetary dust particles in Earth's quasi-satellite resonance}. Icarus
  226, 1550--1558.

\bibitem[{{Lhotka}(2014)}]{Lho14}
{Lhotka}, C., 2014. {Comparitive studies based on the inner, outer and
  equilateral perturbing functions}. Preprint, 1--20.

\bibitem[{{Liou} and {Zook}(1997)}]{Lio97}
{Liou}, J.-C., {Zook}, H.~A., Aug. 1997. {Evolution of Interplanetary Dust
  Particles in Mean Motion Resonances with Planets}. Icarus 128, 354--367.

\bibitem[{{Liou} et~al.(1995){Liou}, {Zook}, and {Jackson}}]{Lio95}
{Liou}, J.-C., {Zook}, H.~A., {Jackson}, A.~A., Jul. 1995. {Radiation pressure,
  Poynting-Robertson drag, and solar wind drag in the restricted three-body
  problem.} Icarus 116, 186--201.

\bibitem[{{Moulton}(1914)}]{Moulton}
{Moulton}, F., 1914. An introduction to celestial mechanics. The Macmillan
  Company, New York.

\bibitem[{{Murray}(1994)}]{Mur94}
{Murray}, C.~D., Dec. 1994. {Dynamical effects of drag in the circular
  restricted three-body problem. 1: Location and stability of the Lagrangian
  equilibrium points}. Icarus 112, 465--484.

\bibitem[{{P{\'a}stor} et~al.(2009{\natexlab{a}}){P{\'a}stor}, {Kla{\v c}ka},
  and {K{\'o}mar}}]{Pas09a}
{P{\'a}stor}, P., {Kla{\v c}ka}, J., {K{\'o}mar}, L., Apr. 2009{\natexlab{a}}.
  {Motion of dust in mean motion resonances with planets}. Celestial Mechanics
  and Dynamical Astronomy 103, 343--364.

\bibitem[{{P{\'a}stor} et~al.(2009{\natexlab{b}}){P{\'a}stor}, {Kla{\v c}ka},
  {Petr{\v z}ala}, and {K{\'o}mar}}]{Pas09b}
{P{\'a}stor}, P., {Kla{\v c}ka}, J., {Petr{\v z}ala}, J., {K{\'o}mar}, L., Jul.
  2009{\natexlab{b}}. {Eccentricity evolution in mean motion resonance and
  non-radial solar wind}. A\&A 501, 367--374.

\bibitem[{{Sicardy} et~al.(1993){Sicardy}, {Beaug\'e}, {Ferraz-Mello},
  {Lazzaro}, and {Roques}}]{Sic93}
{Sicardy}, B., {Beaug\'e}, C., {Ferraz-Mello}, S., {Lazzaro}, D., {Roques}, F.,
  Oct. 1993. {Capture of grains into resonances through Poynting-Robertson
  drag}. Celestial Mechanics and Dynamical Astronomy 57, 373--390.

\bibitem[{{Singh} and {Aminu}(2014)}]{Sin14}
{Singh}, J., {Aminu}, A., Jun. 2014. {Instability of triangular libration
  points in the perturbed photogravitational R3BP with Poynting-Robertson (P-R)
  drag}. Astrophysics \& Space Science 351, 473--482.

\bibitem[{{Stenborg}(2008)}]{Ste08}
{Stenborg}, T.~N., Aug. 2008. {Collinear Lagrange Point Solutions in the
  Circular Restricted Three-Body Problem with Radiation Pressure using
  Fortran}. In: {Argyle}, R.~W., {Bunclark}, P.~S., {Lewis}, J.~R. (Eds.),
  Astronomical Data Analysis Software and Systems XVII. Vol. 394 of
  Astronomical Society of the Pacific Conference Series. pp. 734--737.

\bibitem[{{Stumpff}(1959)}]{Stu59}
{Stumpff}, K., 1959. Himmelsmechanik Band I. Deutscher Verlag der
  Wissenschaften, Berlin.

\bibitem[{{Weidenschilling} and {Jackson}(1993)}]{Wei93}
{Weidenschilling}, S.~J., {Jackson}, A.~A., Aug. 1993. {Orbital resonances and
  Poynting-Robertson drag}. Icarus 104, 244--254.

\end{thebibliography}

\end{document}